\documentclass[%
 reprint,prl,
 amsmath,amssymb,longbibliography,aps,
]{revtex4-1}

\usepackage{natbib}
\usepackage{graphicx}
\usepackage{dcolumn}
\usepackage{bm}

\usepackage{graphicx}
\usepackage{dcolumn}
\usepackage{bm}
\usepackage{hyperref}
\usepackage{xcolor}
\usepackage{amsmath, amssymb, amsfonts}
\usepackage{mathtools}
\usepackage{subfigure}
\usepackage{mathrsfs}
\usepackage{comment}
\usepackage{amsthm}
\usepackage{ dsfont }

\newcommand{\hrho}{\hat{\rho}}

\newcommand{\hrSB}{\hat{\rho}_\mathrm{SB}}
\newcommand{\hrhoS}{\hat{\rho}_\mathrm{S}^\mathrm{ss}}
\newcommand{\ha}{\hat{a}}
\newcommand{\had}{\hat{a}^\dagger}
\newcommand{\hb}{\hat{b}}
\newcommand{\hbd}{\hat{b}^\dagger}

\newcommand{\HS}{\hat{H}_\mathrm{S}}
\newcommand{\HB}{\hat{H}_\mathrm{B}}
\newcommand{\HSU}{\hat{H}_\mathrm{SU}}
\newcommand{\HI}{\hat{H}_\mathrm{I}}
\newcommand{\HIBSU}{\hat{H}_\textrm{I-BSU}}
\newcommand{\vk}{\mathbf{k}}
\newcommand{\omk}{\omega_k}
\newcommand{\kl}{{\mathbf{k}\lambda}}
\newcommand{\klp}{{\mathbf{k}'\lambda'}}
\newcommand{\beq}{\begin{equation}}
\newcommand{\eeq}{\end{equation}}
\newcommand{\beqn}{\begin{eqnarray}}
\newcommand{\eeqn}{\end{eqnarray}}

\begin{document}

\title{Dynamics of a small quantum system open to a bath with thermostat }

\author{Chulan Kwon}
\affiliation{Department of Physics, Myongji University, Yongin, Gyeonggi-Do,
17058,  Korea }
\email{ckwon@mju.ac.kr}
\email{ckwon@mju.ac.kr}
\author{Ju-Yeon Gyhm}
\affiliation{%
Seoul National University, Department of Physics and Astronomy, Center for Theoretical Physics, Seoul 08826, Korea
}
\date{\today}

\begin{abstract}
We investigate dynamics of a small quantum system open to a bath with thermostat. We introduce another bath, called super bath, weakly coupled with the bath to provide it with thermostat, which has either the Lindblad or Redfield type.  We treat the interaction between the system and bath via a rigorous perturbation theory. Due to the thermostat, the bath behaves dissipative and stochastic, for which the usual Born-Markov assumption is not needed. We consider a specific example of a harmonic oscillator system, and a photonic bath in a large container, and a super bath of the Caldeira-Legget oscillators distributed on the inner surface of the container. We use the $P$-representation for the total harmonic system. We derive the reduced time-evolution equation for the system by explicitly finding the correlation between the system and bath beyond the product state, that was not obtainable in the previous theory for the system and bath isolated from environment, and marginalizing bath degrees of freedom. Remarkably, the associated dynamic equation for the system density matrix is of the same form as the Redfield master equation with different coefficients depending on thermostat used. We find steady state does not depend on thermostat, but time-dependent state does, that agrees with common expectation.  We expect to apply our theory to general systems.  Unlike the usual quantum master equations, our reduced dynamics allows investigation for time-dependent protocols and non-equilibrium quantum stochastic dynamics will be investigated in future.      
\end{abstract}
\maketitle


Open quantum systems have usually been studied in the paradigm that a composite system (S) and bath (B), denoted by SB, is isolated from environment~\cite{REDFIELD19651, yoshitaka_kubo1989, Rotter_2015}. Through marginalization of infinite bath degrees of freedom, the Redfield (RF) and Lindbald (LB) quantum master equations have been derived via the Born-Markov assumption~\cite{redfield1957, lindblad, gorini-etal, breuer_petruccione, rivas_heulga}. They have been applied to and many other fields such as quantum optics~\cite{PhysRevLett.68.580, PhysRevLett.110.153601, RevModPhys.88.035002, PhysRevA.105.042405, PhysRevA.95.043826}, transport problems in quantum circuits~\cite{PhysRevE.76.031115, PhysRevA.93.062114, PhysRevLett.121.170601, PhysRevLett.123.180402}, quantum heat engines~\cite{PhysRevA.88.013842, PRXQuantum.2.040328, PhysRevA.90.023819, PhysRevLett.123.240601}. There have been fundamental issues arisen such as complete positivity~\cite{thingnaJCP2012, PhysRevA.101.012103, Davidovic2020completelypositive, nathanPRB2020, beckerPRE2021,tupkaryPRA2022,lee_yeoPRE2022}, local versus global coupling~\cite{levy-kosloff2014,landiRMP2022}, non-equilibrium quantum fluctuations~\cite{talknerPRE2007, campisiPRL2009,  garciaJSTM2012, horowitzNJP2013, horowitzJSP2014, silaevPRE2014, espositoNJP2015}. 

In many experimental situations, a composite SB is open to environment in such a way that B is coupled to a more extensive bath, called a super bath (SU), but S remains apart from SU. We need an extended theory for this new paradigm of tripartite systems, which has been considered in a recent study from the perspective of the relation to the previous paradigm~\cite{tamascelli2018}. 

In this study, we investigate such a tripartite system in detail by dealing with the interaction between B and SU via Born-Markov assumption. Through marginalization of SU degrees of freedom, B is provided with a thermostat, by which B can be thermalized by itself and S is in turn expected to be such via interaction between the two. There arises a fundamental issue on the local versus global coupling scheme for the interaction between SB and SU. From the previous theories, if the interaction between S and B is so weak as to be treated perturbatively, the two coupling schemes are approximately equivalent~\cite{levy-kosloff2014,landiRMP2022}. We observe typical interactions between S and B intrinsically weak enough. We apply the local coupling scheme to marginalize SU degrees of freedom. The global coupling theory requires the knowledge of whole composite eigenstates of SB, which is complicated to find due to infinite B degrees of freedom but can be found for weak interaction limit. We will report the comparison study between the two coupling schemes in a near future. 
 
We consider that B fills an extensive volume $V$ in which S is isolated apart from the boundary $\partial V$ and that B interacts with SU at $\partial V$ which is kept in equilibrium at inverse temperature $\beta$. Let  $\HS$, $\HB$, and  $\HSU$ be respectively the Hamiltonians of the three systems. Let $\HI$ and $\HIBSU$ be respectively interaction Hamiltonians between S and B, and  B and SU. Let $\HB$ and $\HSU$ be  quadratic in their own observables. Let interaction Hamiltonians be bilinear such that $\HI=\hat{S}\otimes\hat{B}$ and $\HIBSU=\hat{B}\otimes\hat{U}$ where $\hat{S}$, $\hat{B}$, $\hat{U}$ denote observables for S, B, SU, respectively. 

We follow the conventional weak coupling theory for the whole interaction $\HI+\HIBSU$~\cite{breuer_petruccione}. Taking the trace over SU states, the cross terms coming from $\HI(t)$ and $\HIBSU(t')$ vanishes. Then, we  obtain the time-evolution equation for the density matrix of SB
\begin{equation}
\dot{\hrho}_\mathrm{SB}=-\frac{i}{\hbar}\left[\HS+\HB+\HI,\hrSB\right]+\mathcal{D}_\mathrm{B}^{\rm X}[\hrSB]
\label{time-evolution}
\end{equation}
where $\mathcal{D}_{\rm B}^{\rm X}$ is the dissipator (superoperator) of only B's operators.  acting on $\hrho_\mathrm{SB}$. B is provided with a thermostat via dissipator of type X, either RF or LB. See the derivation in detail in Sec.~I, Supplementary Material (Suppl.). In the absence of S, the equation leads to the quantum master equation for B, either the RF or LB equation. The appearance of $\mathcal{D}_{\rm B}^{\rm X}$ makes difference from the previous theory for isolated SB.

In the following, we investigate Eq.~(\ref{time-evolution}) via the rigorous perturbation theory for small $\HI$ without using the Born-Markov assumption. B behaves dissipative due to inherent thermostat operated by $\mathcal{D}_{\rm B}^{\rm X}$, while it is assumed to remain in equilibrium in the previous theory for isolated SB. We focus ourselves on a particular case in which S is a single harmonic oscillator located at the center of a cavity of large volume $V$, B is composed of photons in the cavity, and SU composed of  Caldeira-Legett (CL) oscillators~\cite{CALDEIRA1983587} on the inner surface $\partial V$. 

We have $\HB=\sum_{\mathbf{k}\lambda}\hbar\omega_k \hb_{\kl}^\dagger \hb_{\kl}$
for $\omk=|\mathbf{k}|c$ and $\lambda=1, 2$ denote two polarization directions given wave vector $\mathbf{k}$. $\hb_{\kl}$, ($\hbd_\kl$) is  an annihilation (creation) operator for photons with mode given by $\vk$ and $\lambda$. There are nearly infinitely many CL molecules, denoted by $m=1,2,\cdots,M$,  distributed on large $\partial V$. $m$-th oscillator has  angular frequency $\omega_m$, mass $m_m$, charge $q_m$, electric dipole moment $\hat{\mathbf{p}}_m=\mathbf{e}_m q_m\hat{x}_m$ with fixed direction in unit vector $\mathbf{e}_m$ and for instantaneous displacement $\hat{x}_m$ . In terms of ladder operators $\{\hat{c}_m,\hat{c}_m^\dagger\}$, we have $\HSU=\sum_m\hbar\omega_m (\hat{c}_m^\dagger \hat{c}_m+1/2)$ and $\hat{\mathbf{p}}_m =\mathbf{e}_m \kappa_m(\hat{c}_m+\hat{c}_m^\dagger)$ with $\kappa_m=\sqrt{\hbar q_m^2/(2m_m\omega_m)}$. 

We write the second-quantized electric field at $m$-th molecule's position $\mathbf{r}_m$  
\beq
\hat{\mathbf{E}}_m=\sum_{\kl}\mathbf{E}_{\kl,m} \hb_\kl+\mathrm{h.c.}
\eeq
where $\mathbf{E}_{\kl, m}=i\sqrt{2\pi\hbar\omk/V}\mathbf{e}_{\kl}e^{i\mathbf{k}\cdot\mathbf{r}_m}$ and $\mathrm{h.c.}$ denotes the hermitian conjugate of the first term.  $\mathbf{e}_{\mathbf{k}\lambda}$ is a unit vector of polarization. Then, we have interaction Hamiltonian by dipole interaction between photons and CL molecules 
\beq
\HIBSU=-\sum_\omega\sum_{m,a}\hat{E}_{m}^a(\omega) \hat{p}_{m}^a
\eeq 
where $a=1,2,3$ denotes a component of vector. $\hat{E}_m^a(\omega)$ is an eigen-operator of $\hat{E}_m^a$, defined as 
$\hat{E}_{m}^a(\omega)=\sum_{\epsilon_B'-\epsilon_B=\hbar\omega}|\epsilon_B\rangle\langle\epsilon_B | \hat{E}_{m}^a |\epsilon_B'\rangle\langle \epsilon_B'|$
where $|\epsilon_B\rangle$ is an eigenket of $\HB$. We find
\beq
\hat{E}_{m}^a(\omega)={\sum_{\kl}}^\prime \left(\Theta(\omega)E_{\kl,m}^a \hb_\kl +\Theta(-\omega)E_{\kl,m}^{a *} \hb_\kl^\dagger \right)
\eeq
where $\sum^\prime_{\kl}$ denotes the summation restricted for $\omk=|\omega|$ and $\Theta(\omega)$ is the heaviside step function.

Applying the standard weak coupling theory, we find $\mathcal{D}_\mathrm{B}^\mathrm{X}$ expressed in complicated cross terms for different sets of indices, $\{\omega,\kl,m,a\}$ and $\{\omega',\klp,l,b\}$. It can be simplified as follows. From independence of CL molecules, we have $m=l$. On the large $\partial V$, $e^{i(\mathbf{k}-\mathbf{k}')\cdot\mathbf{r}_m}$ is fast oscillating, so we have $\mathbf{k}=\mathbf{k}'$. Requiring a cut-off frequency $\omega_c$ for infinitely many frequencies of CL molecules, $\{\omega_m\}$ forms a nearly continuous distribution. In small window $d\omega$, there are still many values of $\omega_m\in[\omega,\omega+d\omega]$ which are collected from the whole surface so that associated dipole moments are randomly directed. So, we replace $e_m^a e_m^b$ by $\delta_{ab}/3$. Choosing orthogonal polarizations of the electric field for given $\mathbf{k}$, we get $\sum_{a}e^a_{\mathbf{k}\lambda}e^a_{\mathbf{k}\lambda'}=\delta_{\lambda\lambda'}$. As a characteristics of CL model, the spectral density is defined as $J(\omega) =M^{-1}\sum_m q_m^2/(m_m\omega_m)\delta(\omega-\omega_m)$. We present the mathematical procedure in detail in Sec.~II, Suppl.

We present the dissipators of both the RF and LB type in Sec.~II, Suppl. For simplicity, we show the LB type dissipator
\begin{widetext}
\begin{equation}
\mathcal{D}_{\textrm{B}}^{\rm LB}[\rho_{\rm SB}]=\sum_{\mathbf{k}\lambda,\omk<\omega_c}\mu(\omk)
\left[(N(\omk)+1)\left(\hb_{\kl}\hrho_{\rm SB}\hb_{\kl}^\dagger-\frac{1}{2}
\left\{ \hb_{\kl}^\dagger \hb_{\kl},\hrho_{\rm SB}\right\}\right)
+N(\omk)\left(\hb_{\kl}^\dagger\hrho_{\rm SB} \hb_{\kl}-\frac{1}{2}\left\{ \hb_{\kl}\hb_{\kl}^\dagger,\hrho_{\rm SB}\right\}\right)\right]~,
\label{dissipator}
\end{equation}
\end{widetext}
where $\mu(\omk)=\mu_0 \omk J(\omk)$ for $\mu_0=(\pi^2 /3)M/V$.  Writing $M=V^{2/3}/a_0^2$ for mean intramolecular distance $a_0$ of CL molecules, $\mu_0=\pi^2/(3 a_0^2)V^{-1/3}$. $\mu(\omk)\sim V^{-1/3}$ is an important characteristics originating from the boundary interaction between B and SU.  
 
We consider a single harmonic oscillator S of angular frequency $\omega_0$, mass $m$, charge $q$, located at $\mathbf{r}=\mathbf{0}$ inside $V$ and oscillates in $x$ direction. In terms of $\ha$ and $\had$, lowering and raising operators, we have $ \HS=\hbar\omega_0 \left(\ha\had+1/2\right)$ and dipole moment $\hat{p}=\sqrt{\hbar q^2/(2m\omega_0)}(\ha+\ha^\dagger)$.
Then, the diople interaction between S and B gives 
\begin{equation}
\label{interaction}
\HI=-\hat{p} \hat{E}_x(\mathbf{0})=-i\hbar\sum_{\mathbf{k}\lambda} \gamma_{\mathbf{k}\lambda}
(\ha+\had)(\hb_{\kl}-\hb_{\kl}^\dagger)
\end{equation} 
where 
$\gamma_{\mathbf{k}\lambda}=\sqrt{\pi q^2/(m\omega_0V)}\omk^{1/2}e^x_{\mathbf{k}\lambda}$ for $\omk=c|\vk|$.

We regard the composite SB as a collection of harmonic oscillators. Let $\hat{b}_i$ ($\hat{b}_i^\dagger$) for $i=0,1,2,\cdots$ be lowering (raising) operators where $\hat{b}_0=\ha$ and $\hat{b}_i$ for $i\neq 0$ denotes $\hat{b}_{\mathbf{k}\lambda}$. We introduce the coherent state $|z_i\rangle$, the eigenstate of $\hat{b}_i$, given by $e^{-|z_i|^2}/2\sum_n z^n/\sqrt{n!}$. 
Let $|u\rangle=|z,z_1,\cdots,z_i,\cdots\rangle$ be the composite coherent state where $z$ represents eigenvalue of $\hat{b}_0$.
Using the $P$-representation~\cite{glauber1963, sudarshan1963, gardiner2004} in high-dimensional  complex space, we write
$\rho_{\textrm{SB}}=\int d^2 u  P(u,u^*) |u\rangle \langle u |$. 

Using the property: $\hb_i|z_i\rangle=z_i|z_i\rangle\langle z_i|$, $\hb_i^\dagger|z_i\rangle\langle z_i|=\left(\partial/\partial z_i+z_i^*\right)|z_i\rangle\langle z_i|$, we can derive the differential equation for $P(u,u^*)$ from Eq.~(\ref{time-evolution}) to have the form of the Fokker-Planck (FP) equation. Writing $z_i=x_i+iy_i$ for $i>0$ and $z=x+iy$, we find for the LB type thermostat 
\begin{eqnarray}
\lefteqn{\dot{P}=\left[\omega_0\left(-\frac{\partial}{\partial x}y+\frac{\partial}{\partial y}x\right)
+\sum_{i>0}\omega_i\left(-\frac{\partial}{\partial x_i}y_i+\frac{\partial}{\partial y_i}x_i\right)\right.}\nonumber\\
&&-\!\sum_{i>0}\!\gamma_i\!\left[\frac{\partial}{\partial x_i}x-\frac{\partial}{\partial y}y_i\!-\!\frac{1}{4}\left(\frac{\partial^2}{\partial x_i\partial x}
-\frac{\partial^2}{\partial y_i\partial y}\right)\right]
\label{sFP_equation}\\
&&\left.+\sum_{i>0}\frac{\mu_i}{2}\left[
\left(\frac{\partial}{\partial x_i}x_i+\frac{\partial}{\partial y_i}y_i\right)
+\frac{N_i}{2}\left(\frac{\partial^2}{\partial x_i^2}+\frac{\partial^2}{\partial y_i^2}\right)
\right]\right]P~.
\nonumber
\end{eqnarray}
For $i>0$, $\omega_i$, $\gamma_i$, and $\mu_i$ stand for $\omk$, $\gamma_{\mathbf{k}\lambda}$, and $\mu(\omk)$, respectively, and $N_i=(e^{\beta\hbar\omega_i}-1)^{-1}$.  $\sum_{i>0}$ stands for $\sum_{\mathbf{k}\lambda}$, which leads to $V/(2\pi)^3\int d^3\mathbf{k}\sum_{\lambda}$ in continuum limit of $\mathbf{k}$ for large $V$. The expression for the RF type thermostat is given in Sec.~III, Suppl.

Writing $\mathbf{q}=(x,y,x_1,y_1,x_2,y_2,\cdots)^\mathrm{t}$ with subscript t denoting transpose, we get a compact form of the differential equation
\begin{equation}
\label{compact_OU}
\dot{P}(\mathbf{q},t)=\partial_{\mathbf{q}}\cdot(\mathsf{F}\cdot\mathbf{q}+\mathsf{D}\cdot\partial_{\mathbf{q}})P(\mathbf{q},t)
\end{equation} 
where the drift matrix $\mathsf{F}$ and $\mathsf{D}$ are written as
\beq
\mathsf{F}=\left(\begin{array}{cc} \mathsf{F}_{\rm S}& \mathsf{F}_{\rm off}\\ \widetilde{\mathsf{F}}_{\rm off}& \mathsf{F}_{\rm B}\end{array}\right),~\mathsf{D}=\left(\begin{array}{cc} \mathsf{0}& \mathsf{D}_{\rm off}\\  \mathsf{D}_{\rm off}^{\rm t}& \mathsf{D}_{\rm B}\end{array}\right),
\eeq
which is given in detail in Sec.~III, Suppl.
It is an Ornstein-Uhlenbeck (OU) process in infinite dimensions. 
However, $\mathsf{D}$ is not positive definite. Moreover, $P$ can be negative though it is normalizable, so is called quasi-probability distribution~\cite{glauber1963,sudarshan1963}.   

There are two small parameters, $\mu(\omk)\sim V^{-1/3}$ in Eq.~(\ref{dissipator}) and $\gamma_{\mathbf{k}\lambda}\sim V^{-1/2}$ in Eq.~(\ref{interaction}). Both are of dimension of $\omega_0$, so we define dimensionless parameters, $\bar{\mu}(\omk)=\omega_0^{-1}\mu(\omk)$ and $\bar{\gamma}_{\kl}=\omega_0^{-1}\gamma_{\kl}$. 
Intuitively, we expect that the former is much larger than the latter. We need a finer dimensional analysis for the two. We specifically consider the Ohmic spectral density, $J(\omega)=c\omega$, and $c$ is given from $\int_0^{\omega_c}d\omega J(\omega)=c~ \omega_c^2/2=\overline{q_n^2/(m_n\omega_n)}$. Then, we have 
$\bar{\mu}(\omk)=\frac{2\pi(\omk/\omega_0)^2}{3\omega_c/\omega_0}\overline{\left(\frac{q_n^2}{m_n\omega_n}\right)}\frac{1}{\omega_c a_0^2V^{1/3}}$ and 
$\bar{\gamma}_{\mathbf{k}\lambda}=e^x_{\mathbf{k}\lambda}\left[\pi(\nu/ V)(\omk/\omega_0)\right]^{1/2}$ for $\nu=q^2/(m\omega_0^2)$. There are some typical experimental values: electron's charge $=4.8\times 10^{-10}\textrm{esu}$, reduced mass of $\textrm{H}_2$ $=1.7\times 10^{-24}\rm{g}$, typical angular frequency of diatomic molecules $\sim 10^{13}$-$10^{14}\rm{s}^{-1}$, $V\sim 10^3$-$10^6\rm{cm}^3$, and mean intramolecular distance $a_0\sim10^{-8}\rm{cm}$ for $M\sim V^{2/3}/a_0^2$.
Regarding all angular frequencies, $\omega_0,~\omega,~\omega_n$ as having the same order of magnitude, we find $\bar{\gamma}_{\kl},~\bar{\mu}(\omk)\ll 1$, $\bar{\gamma}_{\mathbf{k}\lambda}/{\bar{\mu}(\omega)}\sim\left[m \omega_0^2a_0^4/(q^2V^{1/3})\right]^{1/2}\ll1$. Then, we have an important inequality
\begin{equation}
\bar{\gamma}_{\mathbf{k}\lambda}\ll \bar{\mu}(\omega)\ll 1~.
\label{perturbation}
\end{equation}
According to this inequality, we can perform perturbation expansions in $\bar{\gamma}_{\kl}$ while $\bar{\mu}(\omk)$ is treated rigorously until the end of expansions.  

From the property of the OU process, dissipation (relaxation) to steady state is described by $e^{-\mathsf{F} t}$ accompanied by $e^{-\mathsf{F}^\mathrm{t} t}$. We can find eigenvalues of $\mathsf{F}$ by series expansions in $\bar{\gamma}_{\mathbf{k}\lambda}$ and twice the real parts of eigenvalues give relaxation rates (inverse relaxation times); $\tau_{\rm S}^{-1}$ for S and $\tau_B(\omk)^{-1}$ for photon modes of B. Up to the leading order, $\mathcal{O}(\bar\gamma_{\kl}^2)$, we find that $\tau_{\rm S}^{-1}\simeq\omega_0\nu (\omega_0/c)^3/6$ and $\tau_{\rm B}(\omk)^{-1}\simeq\omega_0\bar{\mu}(\omk)$ depending on frequencies of photon modes; see Sec.~III, Suppl. for derivation. The ratio of the two rates is given as $\tau_{\rm S}^{-1}/\tau_{\rm B}^{-1}\sim \omega_0^3 a_0^2V^{1/3}/c^3\ll 1$. In terms of relaxation times, we have another important inequality 
\begin{equation}
\label{fast and slow}
\tau_{\rm S}\gg\tau_{\rm B}(\omk)\gg \omega_0^{-1}~.
\end{equation}
It implies that B relaxes much faster than S though both relaxations are still very slow compared to the system's internal time scale $\sim\omega_0^{-1}$. Note that this separation of {\it fast and slow} time scales are given intrinsically from the boundary interaction between B and SU. It serves an essential condition in marginalization to average out fast degrees of freedom of B in $P$-distribution. 

The conditional $P$-distribution is defined as the distribution of $\mathbf{q}$ at time $t$ for an initial delta distribution with peak at $\mathbf{q}_0$, formally written as $P(\mathbf{q},t|\mathbf{q}_0,0)\propto e^{-(1/2)(\mathbf{q}-e^{-\mathsf{F}t}\cdot\mathbf{q}_0)^{\rm t}\cdot\mathbf{A}(t)\cdot(\mathbf{q}-e^{-\mathsf{F}t}\cdot\mathbf{q}_0)}$. Writing $\mathbf{q}=(\mathbf{r},\mathbf{R})^{\rm t}$ where $\mathbf{r}^\mathrm{t}=(x,y)$ for S and $\mathbf{R}^{\rm t}=(x_1,y_1, x_2,y_2,\cdots)$ for B, we conjecture 
\begin{equation}
P(\mathbf{r},\mathbf{R},t|\mathbf{r}_0,0)\propto 
e^{-(1/2)\widetilde{\mathbf{R}}^{\rm t}\cdot(\mathsf{A}_{\rm B}+\mathsf{\Gamma})\cdot\widetilde{\mathbf{R}}}
P_{\rm S}(\mathbf{r},t|\mathbf{r}_0,0) 
\label{conjecture1}
\end{equation}
where $\widetilde{\mathbf{R}}=\mathbf{R}-\mathbf{u}$ and $\mathbf{u}=\mathbf{u}(\mathbf{r},\mathbf{r_0})$ is the local equilibrium state of B. Note that $\ln P(\mathbf{r},\mathbf{R},t|\mathbf{r}_0,0)$ should be quadratic in $\mathbf{r}$ and $\mathbf{r_0}$, as characteristic of the OU process. This conjecture reflects that initial memory of $\mathbf{R}_0$ is decayed fast as $e^{-t/\tau_B}$ in relaxation time scale of S. $\mathbf{u}$ is responsible for the correlation between S and B. This conjecture can be confirmed by finding $\mathbf{u}$ and $\mathsf{\Gamma}$ self-consistently. The classical probability density in momentum space in the overdamped limit was conjectured in a similar manner as $\propto e^{-\beta/(2m) (\mathbf{p}-m\mathbf{v})^2}$. It reflects that momentum variables fast reaches a local equilibrium $m\mathbf{v}$. $\mathbf{v}$ was found self-consistently to be the probability current of the derived FP-equation in position space~\cite{ping2007}.

In the absence of interaction, $\mathbf{u}$ and $\mathsf{\Gamma}$  vanish, so the bath $P$-distribution $\propto e^{-(1/2)\mathbf{R}^{\rm t}\cdot\mathsf{A}_{\rm B}\cdot\mathbf{R}}$ corresponds to equilibrium steady state density matrix $\propto e^{-\beta H_{\rm B}}$.  Integrating Eq.~(\ref{conjecture1}) over $\widetilde{\mathbf{R}}$, one get $P_{\rm S}(\mathbf{r},t|\mathbf{r}_0,0)$, which is the conditional $P$-distribution for S given initial state $\mathbf{r}_0$. 

We plug Eq.~(\ref{conjecture1}) into Eq.~(\ref{compact_OU}). We perform perturbation expansion, treating $\mathbf{u}\sim\mathcal{O}(\bar{\gamma}_i)$ and $\mathsf{\Gamma}\sim\mathcal{O}(\bar{\gamma}_i^2)$, which can indeed be confirmed self-consistently.  Integrating the plugged in differential equation over $\widetilde{\mathbf{R}}$, we find the marginal differential equation up to $\mathcal{O}(\bar{\gamma}_i^2)$ 
\begin{equation}
\dot{P}_{\rm S}(\mathbf{r},t|\mathbf{r}_0)=\partial_{\mathbf{r}}\cdot(\mathsf{F}_{\rm S}\cdot\mathbf{r}+\mathsf{F}_{\rm off}\cdot\mathbf{u})P_{\rm S}(\mathbf{r},t|\mathbf{r}_0)~.
\label{marginal_PS}
\end{equation}
and $\mathbf{u}$ is shown to satisfy
\begin{equation}
\dot{\mathbf{u}}=-\mathsf{F}_{\rm B}\cdot\mathbf{u}+\mathbf{r}^{\rm t}\cdot\mathsf{F}_{\rm S}^{\rm t}\cdot\partial_{\mathbf{r}}\mathbf{u}
-\widetilde{\mathsf{F}}_{\rm off}\cdot\mathbf{r}+\widetilde{\mathsf{D}}\cdot\partial_{\mathbf{r}}\ln P_{\rm S}(\mathbf{r},t|\mathbf{r}_0)
\label{u_equation}
\end{equation}
where $\widetilde{\mathsf{D}}=\mathsf{A}_{\rm B}^{-1}\mathsf{F}^{\rm t}_{\rm off}-2\mathsf{D}_{\rm off}^{\rm t}$. The equation for $\mathbf{u}$ is kept up to $\mathcal{O}(\bar{\gamma}_i)$. $\Gamma$ is not needed up to this order. The mathematical steps to derive the marginal equation are given in detail in Sec.~IV, Suppl.

We further conjecture for the conditional $P$-distribution for S as
\begin{equation}
P_{\rm S}(\mathbf{r},t|\mathbf{r}_0,0)=\frac{[{\rm{det}\mathsf{A}}_{\rm S}(t)]^{1/2}}{2\pi}
e^{-(1/2)\widetilde{\mathbf{r}}^{\rm t}\cdot\mathsf{A}_{\rm S}(t)\cdot\widetilde{\mathbf{r}}}
\label{conditional_PS}
\end{equation}
where $\widetilde{\mathbf{r}}=\mathbf{r}-e^{-\mathsf{F}_{\rm eff} t}\cdot\mathbf{r}_0$. It is a formal expression for the OU process~\cite{kwon_ao_thouless,kwonPRE2011}. Using the property that $\mathbf{u}$ is linear in $\mathbf{r}$ and $\mathbf{r}_0$, we write  
$\mathbf{u}=\mathsf{A}_{\rm off}\cdot\mathbf{r}+\mathsf{B}_{\rm off}e^{-\mathsf{F}_{\rm eff} t}\cdot\mathbf{r}_0$.
Through some mathematical steps, we can find $\mathsf{A}_{\rm off}\cdot\mathbf{r},~\mathsf{B}_{\rm off}$ from which $\mathsf{F}_\mathrm{eff}$ and $\mathsf{A}_\mathrm{S}$ can be determined. We 
finally derive the marginal differential equation as
\begin{equation}
\dot{P_S}=\partial_{\mathbf{r}}\cdot\left(\mathsf{F}_{\rm eff}\cdot\mathbf{r}+\mathsf{D}_{\rm eff}\cdot\partial_{\mathbf{r}}\right]P_{\rm S}~.
\label{final_FP}
\end{equation}
where $\mathsf{F}_{\rm eff}$ and $\mathsf{D}_{\rm eff}$ are $2\times 2$ matrices. The derivation is given in detail in Sec.~V, Suppl.

We show 
\begin{equation}
\omega_0^{-1}\mathsf{F}_{\rm eff}=\left(\begin{array}{cc}0&-1\\ 1+ \bar{\nu} f_2&\bar{\nu}f_1\end{array}\right)~,~~\omega_0^{-1}\mathsf{D}_{\rm eff}=\left(\begin{array}{cc}0&\bar{\nu}d_2\\ \bar{\nu}d_2&\bar{\nu}d_1\end{array}\right)~.
\label{simple_matrix}
\end{equation} 
where $\bar{\nu}=\nu(\omega_0/c)^3$ for $\nu= q^2/(m\omega_0^2)$ is small and dimensionless, coming from the perturbation parameter $\bar{\gamma}_{\mathbf{k}\lambda}$. The kernel $\mathsf{A}_{\rm S}$ for the solution can be found from 
\beq\dot{\mathsf{A}}_\mathrm{S}^{-1}=-\mathsf{F}_{\rm eff}\mathsf{A}_\mathrm{S}^{-1}-\mathsf{A}_\mathrm{S}^{-1}\mathsf{F}_{\rm eff}^{\rm t}
+2\mathsf{D}_{\rm eff}.
\eeq. 

Matrix elements, $f_{1,2}$ and $d_{1,2}$, result from sums over all photon modes, hence the integrals over $\mathbf{k}$ in the continuum limit for large $V$. The integral forms can be found in Sec.~VI, Suppl. Finally, changing variable $\xi=\omk/\omega_0$ and taking the limit $\bar{\mu}\to 0$, we find $\omega_c>\omega_0$
\begin{eqnarray}
 f_1&=&\frac{1}{6}R_\mathrm{B}~,\label{f1}\\
 f_2&=&-\frac{1}{3\pi}~ \mathrm{P}\int_0^{\xi_c}\!\!\! \!\!d\xi ~\frac{\xi^4}{\xi^2 -1}~,
 \label{f2}\\
d_1&=&\frac{N_0}{12}R_\mathrm{B}~,\label{d1}\\
d_2&=&\frac{1}{24\pi}~\mathrm{P}\int_0^{\xi_c}\!\!\! \!\!d\xi ~\xi^3\left[\frac{1}{ \xi+1}-\frac{2N(\xi)}{\xi^2-1} \right]~,
\label{d2}
\end{eqnarray}
where $N_0$ is the average number of quanta for S and P denotes the principal value of an integral. 
We have
\beq
R_\mathrm{B}=\left\{\begin{array}{cc} 1~~~;&\textrm{LB-type thermostat}\\
\frac{J_0}{\sqrt{P_0^2+J_0^2}};&\textrm{LF-type thermostat}
\end{array}\right.
\eeq
where $J_0=J(\omega_0)$ and $P_0=\frac{1}{\pi}\mathrm{P}\!\int_0^{\omega_c}\!\!\mathrm{d}\Omega J(\Omega)\frac{2\Omega}{\omega_0^2-\Omega^2}$. Note that $\Omega$ is the variable continued from $\{\omega_m\}$ for the CL oscillators. See the detail derivation in Sec.~VI, Suppl.

It is remarkable that the conventional RF equation for reduced density matrix for isolated SB leads to the same form in $P$-representation as in Eq.~(\ref{final_FP}). However, matrix elements, $f_{1,2}^\mathrm{RF}$ and $d_{1,2}^\mathrm{RF}$, are similar but different from those in Eqs.~(\ref{f1})-(\ref{d2}) obtained for the limit $\bar{\mu}\to 0$. In fact, $f_{1,2}^\mathrm{RF}$ and $d_{1,2}^\mathrm{RF}$ are four times larger than $f_{1,2}$ and $d_{1,2}$ for $R_\mathrm{B}=1$. See the comparison in detail in Sec.~VII, Suppl.

The steady state kernel can be found from $\mathsf{F}_{\rm eff}\mathsf{A}_\mathrm{S}^{-1}+\mathsf{A}\mathrm{S}^{-1}\mathsf{F}_{\rm eff}^{\rm t}
-2\mathsf{D}_{\rm eff}=\mathsf{0}$, found from Sec.~V, Suppl. We get
\beq
\mathsf{A}_\mathrm{S}(\infty)
=\frac{2}{N_0}\left(\begin{array}{cc}
1+C\bar{\nu}&0\\0&1\end{array}\right)~~ \textrm{for $\omega_c\!>\!\omega_0$}.
\eeq
where $C=f_2-(4/N_0)d_2$. The factor $2/N_0$ comes from $f_1/d_1$ in Eqs.~(\ref{f1}) and (\ref{d1}) found for $\omega_c>\omega_0$. In the leading order, it gives the steady state density matrix $\propto e^{-\beta\HS}$. The condition $\omega_c\!>\!\omega_0$ implies that the frequency range of photons should include the system frequency $\omega_0$ for S to be thermalized close to equilibrium. $C$ specifies the correction of $\mathcal{O}(\bar{\nu})$, originating from the second order contribution of $\HI$. It is noteworthy that the steady state kernel be independent of the type of thermostat, shown up to $\mathcal{O}(\bar{\nu})$.   
The steady state density matrix $\hrhoS$ for S can be found inversely from the steady state $P$ distribution $\propto e^{-1/2 \mathbf{r}^\mathrm{t}\cdot \mathsf{A}_\mathrm{S}(\infty)\cdot\mathbf{r}}$. We expand $\hrho_\mathrm{S}^\mathrm{SS}=\hrho^{(0)}+\hrho^{(1)}$, where the first and second terms are of the zeroth and first order in $\bar{\nu}$. We find $\hrho^{(0)}=e^{-\beta\hbar\omega_0(\had\ha+1/2)}/Z_0$ for the partition function $Z_0$, as expected. 
We also find
\beqn
\hrho^{(1)}&=&-\hrho^{(0)}\frac{C\bar{\nu}}{2N_0}e^{-2\beta\hbar\omega_0}\left(\ha\ha+e^{2\beta\hbar\omega_0}\had\had\right.
\nonumber\\
&&\left.+2e^{\beta\hbar\omega_0}\had\ha-2N_0e^{2\beta\hbar\omega_0}\right)~.
\label{correction}
\eeqn
The derivation is given in Sec.~VIII, Suppl. It can be shown not to be equal to the mean force Gibbs (MFG) density matrix $\propto \mathrm{Tr}_\mathrm{B} e^{-\beta(\HS+\HB+\HI}$, which is supposed to the steady state of an isolated SB, hence that of the conventional RF equation~\cite{thingnaJCP2012,lee_yeoPRE2022}. 

The time-dependent state is found to depend on thermostat, as in realistic situations. In particular, the relaxation rate $\tau_S^{-1}$ estimated from $\mathsf{F}_\mathrm{eff}$ is equal to $\omega_0\bar{\nu} f_1$; see Eq.~(\ref{f1}). We consider S to be initially in a pure state with $\rho_S(0)=\sum_{n,m}\rho_{nm}|n\rangle\langle m|$. It corresponds to $P_{\rm S}(z_0, 0)=\sum_{n,m}\rho_{nm}\frac{e^{|z_0|^2}}{\sqrt{n!m!}}\frac{\partial^{n+m}}{\partial(-z_0)^n\partial(-\bar{z}_0)^m}\delta^{(2)}(z_0)$, which is not positive and highly singular, as well known. Then, we can get the marginal $P$ distribution at time $t$ by using Eq.~(\ref{conditional_PS}. As a simple case, we can find the time-dependent $P$-distribution for $\rho_{\rm S}(0)=|1\rangle\langle 1|$, given in Sec.~IX, Suppl. We demonstrate it in Figs.~1 and 2. 
\begin{figure}
 \includegraphics*[width=\columnwidth]{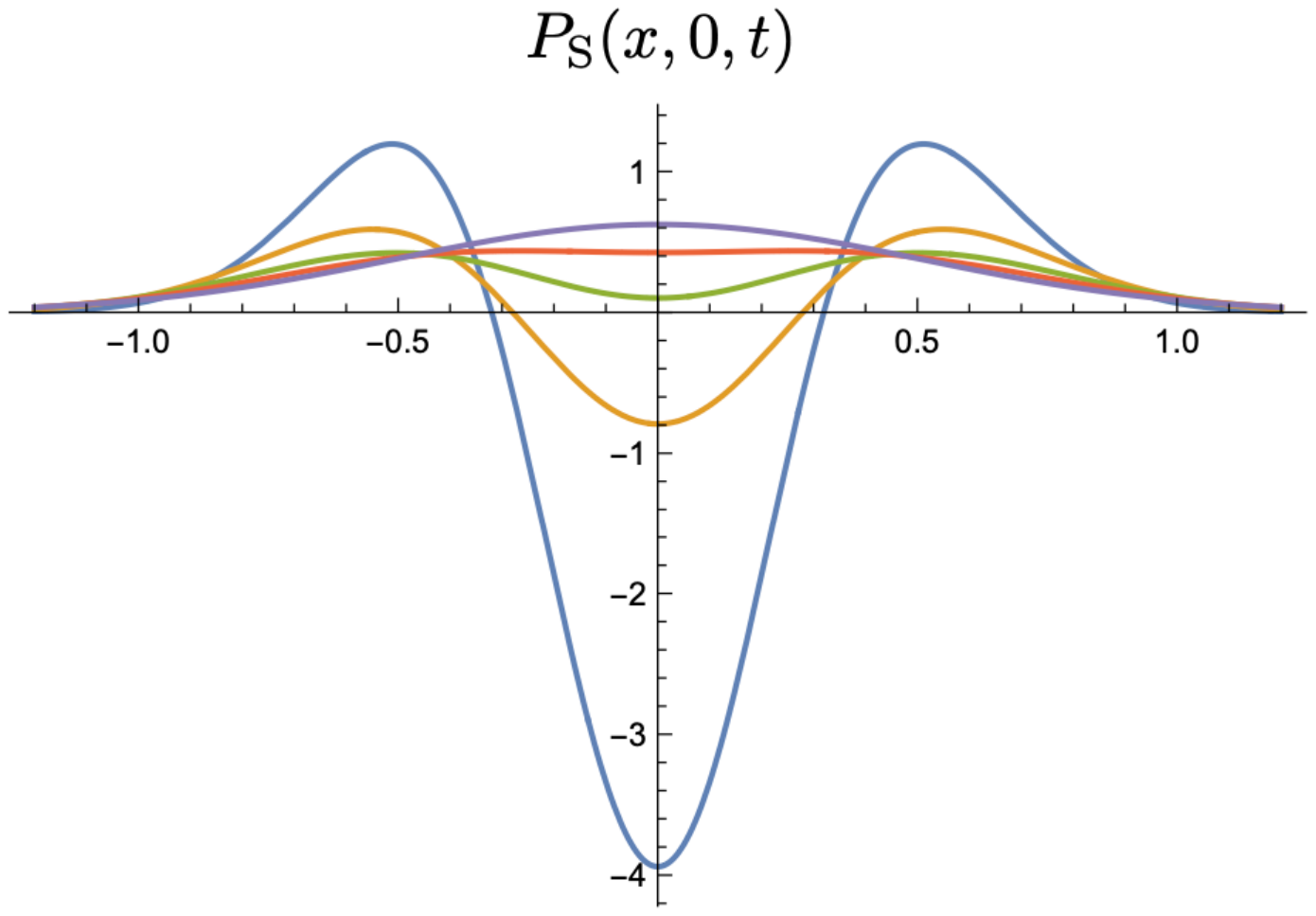}
\caption{$P_{\rm S}(x,y=0,t)$ versus $x$ is plotted for $\bar\nu=0.1$, $\omega_c=2\omega_0$, and $\beta\hbar\omega_0=1$. $\tau_{\rm S}\simeq\ 592\omega_0^{-1}$. The various curves are presented for $\omega_0t=200,~300,~400,~500,~1000$, represented by blue, yellow, green, red, black curves, respectively. As time increases, negative distribution decreases to reach a Gaussian distribution.}
\label{fig1}
\end{figure}
\begin{figure}
 \includegraphics*[width=\columnwidth]{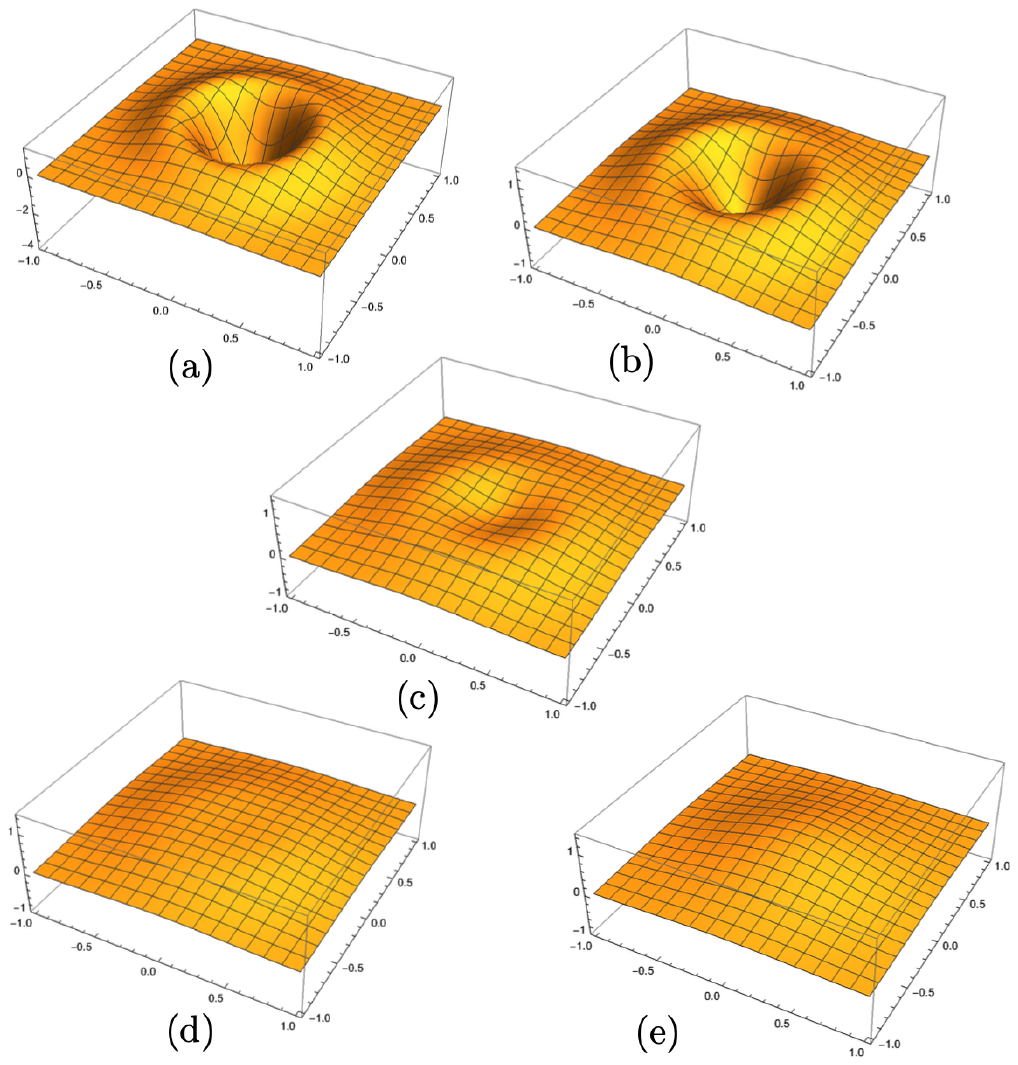}
\caption{$P_{\rm S}(x,y,t)$ is plotted for the same parameters as in Fig.~\ref{fig1}. The panel (a) for $\omega_0t=200$, (b) for $\omega_0t=300$ shows negative distributions inside a valley around the origin. The panel (c) for $\omega_0 t=400$ and  (d) for  $\omega_0 t=500$ shows positive distribution everywhere with shallow valleys. The panel (e) $\omega_0 t=1000$, corresponding to about $2\times \tau_{\rm S}$, shows distribution nearly Gaussian with no valley. }
\label{fig2}
\end{figure}

In conclusion, we derive the marginal differential equation for $P$-distribution for S coupled with B equipped with thermostat. The corresponding equation for reduced density matrix is found to have the same form with different coefficients as the conventional RF equation for isolated SB. Thus, we conclude that the steady state of S in our study is not the MFG state, expected for isolated SB.  We find the steady state does not depend on thermostat, while dynamical process does, which is reasonable in comparison with classical cases where steady state is independent of friction and diffusion coefficients. It is interesting whether or not the MFG state might be recovered by the global coupling scheme to replace $\mathcal{D}_B^\mathrm{X}$. We will generalize our theory to non-harmonic systems such as a two-level system. Since our theory does not require eigen-operators based on eigenstates of the static system, we expect to extend our theory to systems with time-dependent protocols and to investigate non-equilibrium quantum stochastic thermodynamics. 

\begin{acknowledgments}
We are grateful to Professors Junhyun Yeo, Hyunggyu Park, Jaesung Lee, Hyun-Keun Lee, and Jaegon Um for a lot of stimulating discussions and helpful suggestions.
This work was supported by the National Research Foundation of Korea(NRF) grant
No. 2020R1A2C10097611413582110600104 funded by the Korea government(MSIT).

\end{acknowledgments}

\bibliography{bib}
\end{document}



\title{Supplementary material for dynamics of a small quantum system open to a bath with thermostat}

\author{Chulan Kwon}
\affiliation{Department of Physics, Myongji University, Yongin, Gyeonggi-Do,
17058,  Korea}
\email{ckwon@mju.ac.kr}
\author{Ju-Yeon Gyhm}
\affiliation{%
Seoul National University, Department of Physics and Astronomy, Center for Theoretical Physics, Seoul 08826, Korea
}%
\date{\today}
\maketitle

Let  $\hat{H}_{\rm{S}}$, $\hat{H}_{\rm{B}}$, and  $\hat{H}_{\rm{SU}}$ be the Hamiltonians of system (S), bath (B), and super bath (SU), and let $\hat{H}_{\textrm{I}}$ and $\hat{H}_{\textrm{I-BSU}}$ be the interaction Hamiltonians between S and B, B and SU, respectively. We suppose $\hat{H}_{\rm B },~\hat{H}_{\rm SU }$ to be quadratic in their own observables. $\hat{H}_{\rm I}$ ($\hat{H}_{\textrm{I-BSU}}$) is bilinear in each of two observable of S and B (B and SU).

\section{Derivation of time-evolution equation for system and bath with thermostat}
We follow the usual weak coupling theory to treat the interaction $\HIBSU$~\cite{breuer_petruccione}. The density matrix $\hat{\rho}$ for the total system evolves unitarily in time. 
In interaction picture with tilde notation such that $\tilde{O}(t)=e^{it\hat{H}_0/\hbar}\hat{O} e^{-it\hat{H}_0/\hbar}$ for an operator $\hat{O}$ and $\hat{H}_0=\HS+\HB+\HSU$, we have
\beq
\dot{\tilde{\rho}}=-\frac{i}{\hbar}\left[\tilde{H}_{\rm I}(t)+\tilde{H}_\textrm{I-BSU}(t),\tilde{\rho}(t)\right]~.
\label{total}
\eeq
The formal solution is given by
\beq
\tilde{\rho}(t)=\tilde{\rho}(0)-\frac{i}{\hbar}\int_0^t\mathrm{d}t' \left[\tilde{H}_{\rm I}(t')+\tilde{H}_\textrm{I-BSU}(t'),\tilde{\rho}(t')\right]~.
\eeq
The reduced density matrix for the composite system SB is given by $\tilde{\rho}_\mathrm{SB}=\mathrm{Tr}_\mathrm{SU}\tilde{\rho}$. Tracing over SU states, Eq.~(\ref{total}) becomes
\begin{eqnarray}
\label{evolution}
\dot{\tilde{\rho}}_{\rm SB}(t)&=&-\frac{i}{\hbar}\mathrm{Tr}_{\rm SU}\left[\tilde{H}_{\rm I}(t)+\tilde{H}_\textrm{I-BSU}(t),\tilde{\rho}(0)\right]\\
&&\!\!\!-\frac{1}{\hbar^2} \!\!\int_0^t \!\!\mathrm{d}t' \mathrm{Tr}_{\rm SU}\!\!\left[\!\tilde{H}_{\rm I}(t)\!+\!\tilde{H}_\textrm{I-BSU}(t),\right.\nonumber\\
&&\left.\!\left[\tilde{H}_{\rm I}(t')\!+\!\tilde{H}_\textrm{I-BSU}(t'),\tilde{\rho}(t')\!\right]\right]
\nonumber
\end{eqnarray}
We consider an initial condition, $\tilde{\rho}(0)=\tilde{\rho}_{\rm SB}(0)\otimes \tilde{\rho}_{\rm SU}(0)$. As SU remains almost in equilibrium, we can assume 
\beq
\mathrm{Tr}_\mathrm{SU}\left[\tilde{H}_\mathrm{I}(t),\tilde{\rho}_\mathrm{SU}(0)\right]=0~.
\eeq

In the integrand, we use the Born approximation 
\beq
\tilde{\rho}(t')\simeq \tilde{\rho}_{\rm SB}(t')\otimes \tilde{\rho}_\mathrm{SU}^\mathrm{eq}
\eeq 
where $\tilde{\rho}_\mathrm{SU}^\mathrm{eq}=e^{-\beta H_{\rm SU}}/Z_{\rm SU}$ with the partition function $Z_{\rm SU}$.
Then, the cross terms of $\tilde{H}_{\rm I}$ and $\tilde{H}_{\rm I-BSU}$ vanish under the trace over SU. Then, we get 
\begin{eqnarray}
\lefteqn{\dot{\tilde{\rho}}_{\rm SB}(t)=-\frac{i}{\hbar}[\tilde{H}_{\rm I}(t),\tilde{\rho}_{\rm SB}(0)]-\frac{1}{\hbar^2}\int_0^t \!\!\mathrm{d}t' \!\!\left[\tilde{H}_{\rm I}(t),\!\!\left[\tilde{H}_{\rm I}(t'),\tilde{\rho}_\mathrm{SB}(t')  \right]\right]}\nonumber\\
&&-\frac{1}{\hbar^2}\textrm{Tr}_{\rm SU}\!\!\int_0^t \!\!\!\mathrm{d}t' \!\!\left[\tilde{H}_{\textrm{I-BSU}}(t), \left[\tilde{H}_{\textrm{I-BSU}}(t'),\tilde{\rho}_{\rm SB}(t')\otimes\tilde{\rho}_\mathrm{SU}^\mathrm{eq}
~~\right]\right]\nonumber\\
&=&-\frac{i}{\hbar}\left[\tilde{H}_{\rm I}(t),\tilde{\rho}_{\rm SB}(t)\right]\\
&&-\frac{1}{\hbar^2}\textrm{Tr}_{\rm SU} \int_0^t \!\!\!\mathrm{d}t' \!\!\left[\tilde{H}_{\textrm{I-BSU}}(t),\!\!\left[\tilde{H}_\textrm{I-BSU}(t'),\tilde{\rho}_{\rm SB}(t')\otimes\tilde{\rho}_\mathrm{SU}^\mathrm{eq}\right]\right]\nonumber
\end{eqnarray}
The correlation of SU observables in $\tilde{H}_{\textrm{I-BSU}}(t)\tilde{H}_\textrm{I-BSU}(t')$ in the integrand of the above equation is expected to change very fast compared to relatively slowly varying $\tilde{\rho}_{\rm SB}(t')$, so we can replace $\tilde{\rho}_{\rm SB}(t')$ with $\tilde{\rho}_{\rm SB}(t)$. It is the essence of 
the Markov approximation that is a key scheme along with the Born approximation in the weak coupling theory. 
We can also extend the lower limit $t'=0$ to $-\infty$. 

Then, we get
\beqn
\label{prestep1}
\lefteqn{\dot{\tilde{\rho}}_{\rm SB}(t)=-\frac{i}{\hbar}\left[\tilde{H}_{\rm I}(t),\tilde{\rho}_{\rm SB}(t)\right]}\\
&&\!-\!\frac{1}{\hbar^2}\textrm{Tr}_{\rm SU}\! \!\!\int_0^\infty\!\!\!\!\!\!\mathrm{d}s\!\left[\tilde{H}_{\textrm{I-BSU}}(t),\!\!\left[\tilde{H}_\textrm{I-BSU}(t\!-\!s),\tilde{\rho}_{\rm SB}(t)\otimes\tilde{\rho}_\mathrm{SU}^\mathrm{eq}\right]\right]\nonumber\\
&=&-\frac{i}{\hbar}\left[\tilde{H}_{\rm I}(t),\tilde{\rho}_{\rm SB}(t)\right]+\tilde{\mathcal{D}}_\mathrm{B}^\mathrm{X}[\tilde{\rho}_{\rm SB}(t)]
\nonumber
\eeqn
First, note that the first term with the commutator is exact and the second is given up to $\mathcal{\HIBSU}$. $\tilde{\mathcal{D}}_\mathrm{B}^\mathrm{X}$ is a superoperator, called dissipator in the interaction picture,  defined as the traced integral in the above line. The superscript denotes the type; either Redfield (RF) or Lindblad (LB), which will be derived in the next section.  It is given by B observables only, so gives rise to stochastic and dissipative nature to B. It is a theoretically designed thermostat for B provided by SU, comparable to Langevin thermostat in classical cases. This thermal nature is transferred to S via interaction $\HI$.  In the Schr\"{o}dinger picture, we have 
\begin{equation}
\dot{\hat{\rho}}_{\rm SB}=-\frac{i}{\hbar}\left[\hat{H}_{\rm S}+\hat{H}_{\rm B}+\hat{H}_{\rm I},\hat{\rho}_{\rm SB}\right]+\mathcal{D}_{\rm B}^X[\hat{\rho}_{\rm SB}]
\label{prestep2}
\end{equation}
It is the starting point in our study, Eq.~(1) in text. 

\section{Derivation of the dissipator for photonic bath and Caldeira-Legget super bath}

The dissipator of B in interaction picture is defined from Eq.~(\ref{prestep1})
\beqn
\label{dissipator_X}
\lefteqn{\tilde{\mathcal{D}}_\mathrm{B}^\mathrm{X}[\tilde{\rho}_{\rm SB}(t)]=}\\
&&\!-\!\frac{1}{\hbar^2}\textrm{Tr}_{\rm SU}\! \!\!\int_0^\infty\!\!\!\!\!\!\mathrm{d}s\!\left[\tilde{H}_{\textrm{I-BSU}}(t),\!\!\left[\tilde{H}_\textrm{I-BSU}(t\!-\!s),\tilde{\rho}_{\rm SB}(t)\otimes\tilde{\rho}_\mathrm{SU}^\mathrm{eq}\right]\right]~.\nonumber
\eeqn
We will derive it for the photonic bath in a cavity of large volume $V$ and the super bath composed of  Caldeira-Legget (CL) molecules distributed on the inner surface $\partial V$ of the cavity. 

For photonic bath, $\HB=\sum_{\kl}\hbar\omk\hb_\kl^\dagger b_\kl$
for $\omk=|\vk|c$ and polarization index $\lambda=1, 2$. For the Caldeira-Legget (CL) molecules, denoted by $m=1,\cdots, M$ for infinitely large $M$, $\hat{H}_{\rm SU}=\sum_m\hbar\omega_m (\hc_m^\dagger \hc_m+1/2)$ with dipole moment $\hat{\mathbf{p}}_m =\mathbf{e}_m \kappa_m(\hc_m+\hc_m^\dagger)$ with $\kappa_m=\sqrt{\hbar q_m^2/(2m_m\omega_m)}$ for charge $q_m$, mass $m_m$, angular frequency $\omega_m$, and unit vector $\mathbf{e}_m$ of oscillation. We suppose that photons of B interacts with CL dipoles via dipole interaction given by 
$\hat{H}_\textrm{I-BSU}=-\sum_{m} \hat{\mathbf{E}}_m\cdot\hat{\mathbf{p}}_m$. $\hat{\mathbf{E}}_m$ is the electric field at the position $\mathbf{r}=\mathbf{r}_m$ of $m$-th dipole
\begin{eqnarray}
\hat{\mathbf{E}}_m&=&i\sum_{\kl}\sqrt{\frac{2\pi\hbar\omk}{V}}\mathbf{e}_\kl (\hb_{\kl} e^{i\vk\cdot\mathbf{r}_m}-\hb_{\kl}^\dagger e^{-i\vk\cdot\mathbf{r}_m})\nonumber\\
&=&\sum_{\kl}(\mathbf{E}_{\kl,m}\hb_{\kl}+\mathbf{E}_{\kl,m}^* \hb_\kl^\dagger)
\end{eqnarray}
where $\mathbf{E}_{\kl,m}=i\sqrt{2\pi\hbar\omk/V}\mathbf{e}_\kl e^{i\vk\cdot\mathbf{r}_m}$
and $\mathbf{e}_{\mathbf{k}\lambda}$ is a unit polarization vector. 

We write $\hat{H}_\textrm{I-BSU}=-\sum_{m,a} \hat{E}_m^a \hat{p}_m^a$ for superscript $a=1,2,3$ denoting the component of vector. Associated with this interaction form, eigenoperator is defined for B as
\begin{equation}
\hat{E}_{m}^a(\omega)=\sum_{\epsilon_B'-\epsilon_B=\hbar\omega}|\epsilon_B\rangle\langle\epsilon_B | \hat{E}_{m}^a |\epsilon_B'\rangle\langle \epsilon_B'|
\label{eigen-operator}
\end{equation}
where $|\epsilon_B\rangle$ is an eigenket of $\HB$.
We find
\begin{equation}
\hat{E}_{m}^a(\omega)=\left\{\begin{array}{cc}\sum_{\kl}^{\prime} E_{\kl,m}^{a} \hb_\kl & \textrm{for $\omega>0$}\\
\sum_{\kl }^\prime E_{\kl,m}^{a *} \hb_\kl^\dagger & \textrm{for $\omega<0$}\\
\end{array}\right.
\end{equation}
where $\sum_{\kl}^{\prime}$ denotes the summation restricted for $|\vk|=|\omega|/c$. Eigenoperators are found to have 
properties
\begin{eqnarray}
\label{superop_prop}
 \hat{E}_{m}^{a\dagger}(\omega)&=&\hat{E}_{m}^{a}(-\omega),\nonumber\\
 e^{c \HB}\hat{E}_{m}^a(\omega)e^{-c \HB}&=&e^{-c\hbar\omega}\hat{E}_{m}^a(\omega),\\
 e^{c^* \HB}\hat{E}_{m}^{a\dagger}(\omega) e^{-c^* \HB}&=&e^{-c^*\hbar\omega}\hat{E}_{m}^{a\dagger}(\omega)
 \nonumber
\end{eqnarray}
where $c=it/\hbar$ is for interaction picture and $c=\beta$ is useful. 
Then, we can write 
\beq
\HIBSU=-\sum_{\omega}\sum_{m,a}\hat{E}_{m}^a(\omega)\hat{p}_m^a=-\sum_{\omega'}\sum_{l,b}\hat{E}_{l}^b(\omega')^\dagger\hat{p}_l^b\\
\label{HIBSU}
\eeq
where we only change dummy indices for the last expression. Using Eq.~(\ref{superop_prop}), we write
\beq
\label{int_form}
\begin{split}
\tilde{H}_\textrm{I-BSU}(t\!-\!s)=\!-\!\sum_{\omega}\!\sum_{m,a}\!e^{-i\omega(t-s)}\hat{E}_{m}^a(\omega)\tilde{p}_m^a(t\!-\!s)\\
\tilde{H}_\textrm{I-BSU}(t)=\!-\!\sum_{\omega'}\sum_{l,b}\!e^{i\omega't}\hat{E}_{l}^b(\omega)^\dagger\tilde{p}_l^b(t)~~~~~~~~~~
\end{split}
\eeq
where dipole moment in interaction picture is given such that 
\beq
\tilde{p}_m^a(t\!-\!s)=e_m^a\kappa_m(e^{-i\omega_m (t\!-\!s)}\hat{c}_m+e^{i\omega_m (t\!-\!s)}\hat{c}_m^\dagger)~.
\eeq

Pugging Eq.~(\ref{int_form}) into Eq.~(\ref{dissipator_X}), we find the RF-type dissipator of B in interaction picture
\beqn
\tilde{\mathcal{D}}_\mathrm{B}^\mathrm{RF}[\tilde{\rho}_{\rm SB}]&=&\frac{1}{\hbar^2}\sum_{\omega,\omega'}\sum_{m,a}\sum_{l,b}e^{-i(\omega-\omega')t}\Gamma_{ml}^{ab}(\omega)\nonumber\\
&&\times\left(\hat{E}_m^a(\omega)\tilde{\rho}_{\rm SB}\hat{E}_{l}^{b\dagger}(\omega') -\hat{E}_l^{b\dagger}(\omega')\hat{E}_{m}^{a}(\omega) \tilde{\rho}_{\rm SB}\right)~,
\nonumber\\
&&+\mathrm{h.c.}
\label{dissipator_RF}
\eeqn
where $\Gamma_{ml}^{ab}(\omega)$ is found as
\begin{eqnarray}
\lefteqn{\Gamma_{ml}^{ab}(\omega)= \kappa_m\kappa_l e_m^a e_l^b \lim_{\epsilon\to 0^+}\int_0^\infty \mathrm{d}s e^{i\omega s-\epsilon s}}\nonumber\\
&&\times\mathrm{Tr}_{\rm B}\Big[(\hc_m e^{-i\omega_m s}+\hc_m^\dagger e^{i\omega_m s})(\hc_l+\hc_l^\dagger)e^{-\beta \HSU}/Z_\mathrm{SU}\Big]\nonumber\\
&=&\kappa_m^2 e_m^a e_m^b\delta_{ml}\left[ (N(\omega_m)+1)\left(\pi\delta(\omega-\omega_m)+i\mathrm{P}\frac{1}{\omega-\omega_m}\right)\right.\nonumber\\
&&\left. +N(\omega_m)\left(\pi \delta(\omega+\omega_m)+i\mathrm{P}\frac{1}{\omega+\omega_m}\right)\right]\nonumber\\
&=&\delta_{ml}\kappa_m^2e_m^ae_m^b\Gamma_{m}(\omega)
\label{Gamma}
\end{eqnarray}
where $\Gamma_m(\omega)$ is equal to the terms in the square bracket. We use $\mathrm{Tr}_{\rm B} \hc_m^\dagger \hc_l e^{-\beta \HB}/Z_\mathrm{B}=\delta_{ml} N(\omega_m)$ with  $N(\omega_m)=(e^{\beta\hbar\omega_m}-1)^{-1}$ and also use $\int_0^\infty \mathrm{d}s e^{i(\omega-\omega_m) s-\epsilon s}=\frac{1}{\epsilon-i(\omega-\omega_m)}=\pi\delta(\omega-\omega_m)+i \mathrm{P}\frac{1}{\omega-\omega_m}$ for $\mathrm{P}$ denoting the principal value. 

In the Schr\"{o}dinger picture, we have $\mathcal{D}_\mathrm{B}^\mathrm{X}=e^{-it\HB/\hbar}\tilde{\mathcal{D}}_\mathrm{B}^\mathrm{X}e^{it\HB/\hbar}$. We find
\beqn
\label{L_RF}
\lefteqn{\mathcal{D}_\mathrm{B}^\mathrm{RF}[\hrho_{\rm SB}]=\frac{1}{\hbar^2}\sum_{\omega,\omega'}\sum_{m,a,b}\kappa_m^2e_m^a e_m^b\Gamma_{m}(\omega)}\\
&&\times\left(\hat{E}_m^a(\omega)\hrho_{\rm SB}\hat{E}_{l}^{b\dagger}(\omega') -\hat{E}_l^{b\dagger}(\omega')\hat{E}_{m}^{a}(\omega) \hrho_{\rm SB}\right)+\mathrm{h.c.}\nonumber
\eeqn
In random wave approximation, fast oscillating factor $e^{i(\omega-\omega')}$ is set to zero except for $\omega=\omega'$ in Eq.~(\ref{dissipator_RF}). Then, we find the LB-type dissipator in the Schr\"{o}dinger picture
\beqn
\label{L_LB}
\lefteqn{\mathcal{D}_\mathrm{B}^\mathrm{LB}[\tilde{\rho}_{\rm SB}]=\frac{1}{\hbar^2}\sum_{\omega}\sum_{m,a,b}\kappa_m^2e_m^a e_m^b\Gamma_{m}(\omega)}\\
&&\times\left(\hat{E}_m^a(\omega)\hrho_{\rm SB}\hat{E}_{m}^{b\dagger}(\omega) -\hat{E}_m^{b\dagger}(\omega)\hat{E}_{m}^{a}(\omega) \hrho_{\rm SB}\right)+\mathrm{h.c.}
\nonumber
\eeqn
Thus, one can reduce $\mathcal{D}_\mathrm{B}^\mathrm{RF}$ to $\mathcal{D}_\mathrm{B}^\mathrm{LB}$ by taking $\omega=\omega'$. 

These expressions for dissipators are still complicated as there are coupled terms with different sets of indices, $\{\omega,\mathbf{k},\lambda,a\}$ and  
$\{\omega',\mathbf{k}',\lambda',b\}$. We can simplify them more as follows. $e^{i(\vk\mp\vk')\cdot\mathbf{r}_m}$ from $\hat{E}_m^a(\omega)\hat{E}_m^b(\omega')$ is fast oscillating as $|\mathbf{r}_m|$ is very large on $\partial V$, so we get $\vk=\pm \vk'$ and $|\omega|=|\omega'|$. There are infinitely many CL molecules on large $\partial V$. We consider a cut-off frequency $\omega_c$ with all $\omega_m<\omega_c$, that is physically reasonable. Then, $\{\omega_n\}$ forms a continuous distribution. In small window $[\omega,\omega+d\omega]$, there are still many frequencies associated with randomly oriented dipole moments. Therefore, we can replace $e_m^a e_m^b$ by $\delta_{ab}/3$. From $\sum_a\hat{E}_m^a(\omega)\hat{E}_m^a(\pm \omega)$, we get $\sum_a e_{\kl}^ae_{\kl'}^a=\delta_{\lambda\lambda'}$ for the choice of orthogonal polarizations.

As a characteristic of CL molecules, the spectral density $J(\omega)$ is defined as 
\beq
J(\omega)=\frac{1}{M}\sum_{m=1}^M \frac{q_m^2}{m_m\omega_m}\delta(\omega-\omega_m)~,
\eeq
where $\kappa_m^2=\hbar q_m^2/(2m_m \omega_m)$ is used. Then, we find 
\beqn
\lefteqn{\frac{1}{M}\sum_{m=1}^M \frac{q_m^2}{m_m\omega_m}\Gamma_m(\omega)}\nonumber\\
&=&\pi\Theta(\omega)(N(\omega)+1)J(\omega)+\pi\Theta(-\omega) N(|\omega|)J(|\omega|)\nonumber\\
&&+i\mathrm{P}\!\!\int_0^{\omega_c}\!\!\!\!\mathrm{d}\Omega J(\Omega)\left(\frac{N(\Omega)+1}{\omega-\Omega}+\frac{N(\Omega)}{\omega+\Omega}\right)~.
\eeqn
where $\Theta(\omega)=1$ for $\omega>0$, otherwise $\Theta(\omega)=0$.

We can simplify $\mathcal{D}_\mathrm{B}^\mathrm{RF}$ in Eq.~(\ref{L_RF}) 
\beqn
\lefteqn{\mathcal{D}_\mathrm{B}^\mathrm{RF}[\hrho_{\rm SB}]={\sum_{\kl}}\sum_{\omega,\omega'=\pm\omk}{\sum_{\kl}} \frac{\pi^2 M}{3V}
\left(\!\frac{1}{2}\Gamma(\omega)+iS(\omega)\!\right) }\\
&&\times\left(\hat{B}_\kl(\omega)\hrho_{\rm SB}\hat{B}_\kl^\dagger(\omega') -\hat{B}_\kl^\dagger(\omega')\hat{B}_\kl(\omega) \hrho_{\rm SB}\right)+\mathrm{h.c.}\nonumber
\eeqn
where
\beq
\hat{B}_{\kl}(\omega)=\Theta(\omega)\hat{b}_{\kl}+\Theta(-\omega)\hat{b}_{\kl}^\dagger~.
\eeq
and
\beqn
\frac{1}{2}\Gamma(\omega)&=& |\omega| |N(\omega)+1|J(|\omega|)\\
S(\omega)&=&\frac{1}{\pi}|\omega|\mathrm{P}\!\!\!\int_0^{\omega_c}\mathrm{d}\Omega\left(\frac{N(\Omega)+1}{\omega-\Omega}+\frac{N(\Omega)}{\omega+\Omega}\right)
\eeqn
where $N(-\omega)+1=-N(\omega)$ is used. 

After some algebra, we find a lengthy expression for $\mathcal{D}_\mathrm{B}^\mathrm{RF}$ as follows:
\beq
\label{DissB_RF}
\mathcal{D}_\mathrm{B}^\mathrm{RF}[\hrho_{\rm SB}]=
\sum_{\kl}\mu_0\omk\left(\mathcal{D}_\kl^\mathrm{LB}[\hrho_{\rm SB}]-i[\hat{H}_\kl,\hrho_{\rm SB}]+\mathcal{D}_\kl^\mathrm{add}[\hrho_{\rm SB}]\right)
\eeq
for $\mu_0=2\pi^2 M/(3V)$. Each term is given as 
\beqn
\lefteqn{\mathcal{D}_\kl^\mathrm{LB}[\hrho_{\rm SB}]=}\nonumber\\
&&\!J(\omk)(N(\omk)\!+\!1)\!\!\left( \!\hat{b}_{\kl}\hat{\rho}_{\rm SB}\hat{b}_{\kl}^{\dagger}\! \!-\frac{1}{2}\!\left\{\hat{b}_{\kl}^{\dagger} \hat{b}_{\kl} , \hat{\rho}_{\rm SB}\right\} \!\!\right)\nonumber\\
&&+J(\omk)N(\omk)\!\!\left(\! \hat{b}_{\kl}^{\dagger}\hat{\rho}_{\rm SB}\hat{b}_{\kl}\! -\!\frac{1}{2}\!\left\{\hat{b}_{\kl} \hat{b}_{\kl}^{\dagger} , \hat{\rho}_{\rm SB}\right\} \!\!\right)~~~~
\eeqn
and
\beq
\hat{H}_\kl=\frac{1}{2\pi}S_+(\omk)\left(\hat{b}_{\kl}^\dagger\hat{b}_{\kl}+\frac{1}{2}(\hat{b}_{\kl}\hat{b}_{\kl}+\hat{b}_{\kl}^\dagger\hat{b}_{\kl}^\dagger)\right)
\eeq
and
\beqn
\lefteqn{\mathcal{D}_\kl^\mathrm{add}[\hrho_{\rm SB}]=\frac{1}{2}J(\omk)(N(\omk)+1)\Big( \hat{b}_\kl \hrho_{\rm SB} \hat{b}_\kl+\hat{b}_\kl^\dagger \hrho_{\rm SB}\hat{b}_\kl^\dagger }\nonumber\\
&&-\hat{b}_\kl\hat{b}_\kl\hrho_{\rm SB}-\hrho_{\rm SB}\hat{b}_\kl^\dagger \hat{b}_\kl^\dagger\Big) 
+\frac{1}{2}J(\omega)N(\omega)\Big( \hat{b}_\kl \hrho_{\rm SB} \hat{b}_\kl\nonumber\\
&&~~~~~~+\hat{b}_\kl^\dagger \hrho_{\rm SB}\hat{b}_\kl^\dagger-\hat{b}_\kl^\dagger\hat{b}_\kl^\dagger\hrho_{\rm SB}-\hrho_{\rm SB}\hat{b}_\kl \hat{b}_\kl\Big) \nonumber\\
&&+\frac{i}{2\pi}S_{-}(\omk)\Big( \hat{b}_\kl \hrho_{\rm SB} \hat{b}_\kl-\hat{b}_\kl^\dagger \hrho_{\rm SB}\hat{b}_\kl^\dagger \nonumber\\
&&~~~~~~~-\frac{1}{2}\left\{\hat{b}_\kl\hat{b}_\kl-\hat{b}_\kl^\dagger \hat{b}_\kl^\dagger ,\hat{\rho}_{\rm SB}\right\}\Big)
\eeqn
where 
\beqn
S_{+}(\omk)&=&\mathrm{P}\!\int_0^{\omega_c}\!\!\!\!\!\mathrm{d}\Omega J(\Omega)\frac{2\Omega}{\omk^2-\Omega^2}\\
S_{-}(\omk)&=&\mathrm{P}\!\int_0^{\omega_c}\!\!\!\!\!\mathrm{d}\Omega J(\Omega)\frac{2\omk(2N(\Omega)+1)}{\omk^2-\Omega^2}~.
\eeqn
$\{A,B\}$ is the anti-commutator defined as $AB+BA$ .

$\mathcal{D}_\mathrm{B}^\mathrm{LB}$ is given from the first part of $\mathcal{D}_\mathrm{B}^\mathrm{LF}$ as
\beq
\label{DissB_LB}
\mathcal{D}_\mathrm{B}^\mathrm{LB}[\hrho_{\rm SB}]=\sum_{\kl}\mu_0\omk\mathcal{D}_\kl^\mathrm{LB}[\hrho_{\rm SB}]
\eeq
which is presented in the text by using
\beq
\mu(\omk)=\mu_0 \omk J(\omk)
\eeq 

\section{P-representation for harmonic system and bath}
We consider  
$\HS=\hbar\omega_0 \left(aa^\dagger+\frac{1}{2}\right)$ and the dipole meoment in $x$ direction $p_x=q\sqrt{\frac{\hbar}{2m\omega_0}}(a+a^\dagger)$. Dipole interaction between S and B gives
\beqn
\label{H_I}
\HI&=&-i\sum_{\kl} \sqrt{\frac{\pi \hbar^2 q^2 \omk }{m\omega_0 V}}\mathbf{e}_{\kl}\cdot \hat{\mathbf{x}}
(a+a^\dagger)(b_{\mathbf{k}\lambda}-b_{\mathbf{k}\lambda}^\dagger)\nonumber\\
&=&-i\hbar \sum_{\kl}\gamma_{\kl}(a+a^\dagger)(b_{\kl}-b_{\kl}^\dagger)~.
\eeqn
The intensity of $\HI$ is given by  
\beq
\gamma_\kl=\sqrt{\frac{\pi q^2 \omk }{m\omega_0 V}}e^x_{\kl}=\omega_0 \sqrt{\frac{\pi\nu}{V}\frac{\omk}{\omega_0}} e^x_{\kl}
\label{gamma}
\eeq
where $\nu= q^2/(m\omega_0^2)$ has the dimension of volume. Note that $\gamma_{\kl}$ has the same dimension as $\omega_0$, so does $\mu(\omk)$. We define dimensionless parameters
\beq
\bar{\gamma}_{\kl}=\omega_0^{-1}\gamma_\kl,~\bar{\mu}=\omega_0^{-1}\mu(\omk)
\eeq
In Eq.~(10) in the text, we get for typical experimental parameters 
\beq
\begin{split}
\bar{\gamma}_{\kl},~ \bar{\mu}(\omk)\ll 1\\
\bar{\gamma}_{\kl}\ll \bar{\mu}(\omk)\ll \omega_0
\end{split}
\eeq 
We will can carry out the perturbation theory for small $\bar{\gamma}_{\kl}$ treating $\bar{\mu}(\omk)$ rigorously and take the limit $\bar{\mu}(\omk)\to 0$ in the end of calculation. 

We investigate 
\beq
\dot{\hat{\rho}}_\mathrm{SB}=-\frac{i}{\hbar}\left[ \hbar\omega_0 a^\dagger a+\HI, ~\hat{\rho}_\mathrm{SB}\right]+\mathcal{D}_\mathrm{B}^\mathrm{LB}[\hat{\rho}_\mathrm{SB}]~.
\label{time-evolution}
\eeq
where we use the LB-type dissipator for simplicity. Generalization to the RF-type dissipator will be made. 
S and B are harmonic systems and can be treated as the same footing. Let $\hb_i$ ($\hbd_i$) for $i=0,1,2,\cdots$ be lowering (raising) operators of the total system of S and B where $\hb_0$ denote $\ha$ and $\hb_i$ for $i>0$ denote $\hb_{\mathbf{k}\lambda}$. Then, we introduce the coherent states that are eigenstates of $\hb_i$ with properties~\cite{gardiner2004}
\beq
\begin{split}
|z_i\rangle=e^{-|z_i|^/2}\sum\frac{z^m}{\sqrt{m!}}|m\rangle\langle m|~~~~~~~~~~~~~~~~~\\
\hb_i|z_i\rangle\langle z_i|=z_i|z_i\rangle\langle z_i|,~
\hbd_i|\rangle z_i\langle z_i|=\left(\frac{\partial}{\partial z_i}+z_i^*\right)|z_i\rangle\langle z_i|~\\
\end{split}
\eeq
For $|u\rangle=|z_0,z_1,\cdots,z_i,\cdots\rangle$,  $P$-distribution is defined such that  $\hat{\rho}_\mathrm{SB}=\int du du^* P(u,u^*) |u\rangle \langle u|$~\cite{glauber1963,sudarshan1963}.

$P(u,u^*,t)$ associated with Eq.~(\ref{time-evolution}) is found to satisfy the Fokker-Planck (FP) type equation. Writing $z_i=x_i+iy_i$, we have in the text
\begin{eqnarray}
\label{SFP-eq}
\lefteqn{\dot{P}=\left[\omega_0\left(-\frac{\partial}{\partial x_0}y_0+\frac{\partial}{\partial y_0}x_0\right)
+\sum_{i>0}\omega_i\left(-\frac{\partial}{\partial x_i}y_i+\frac{\partial}{\partial y_i}x_i\right)\right.}\nonumber\\
&&-\!\sum_{i>0}\!\gamma_i\!\left[\frac{\partial}{\partial x_i}x_0-\frac{\partial}{\partial y_0}y_i\!-\!\frac{1}{4}\left(\frac{\partial^2}{\partial x_i\partial x_0}
-\frac{\partial^2}{\partial y_i\partial y_0}\right)\right]\\
&&\left.+\sum_{i>0}\frac{\mu_i}{2}\left[
\left(\frac{\partial}{\partial x_i}x_i+\frac{\partial}{\partial y_i}y_i\right)
+\frac{N_i}{2}\left(\frac{\partial^2}{\partial x_i^2}+\frac{\partial^2}{\partial y_i^2}\right)
\right]\right]P~.
\nonumber
\end{eqnarray}
For $i>0$, $\omega_i$, $\gamma_i$, and $\mu_i$ stand for $\omk$, $\gamma_\kl$, and $\mu(\omk)$, respectively, and  $N_i=(e^{\beta\hbar\omega_i}-1)^{-1}$.

We write $\mathbf{q}=(x_0,y_0,x_1,y_1,x_2,y_2,\cdots,x_N,y_N)^\mathrm{t}$ where the superscript t denoting the transposition of matrix. Then, Eq.~(\ref{SFP-eq}) is written compactly as 
\begin{equation}
\dot{P}(\mathbf{q},t)=\partial_{\mathbf{r}}\cdot(\mathsf{F}\cdot\mathbf{q}+\mathsf{D}\cdot\partial_{\mathbf{r}})P(\mathbf{q},t)
\label{OU_process}
\end{equation}
where the drift and diffusion matrices have block structure such that
\begin{equation}
\mathsf{F}=\left(\begin{array}{cc}\mathsf{F}_{\rm S}&\mathsf{F}_{\rm off}\\
\tilde{\mathsf{F}}_{\rm off}&\mathsf{F}_{\rm B}\end{array}\right)~,~~
\mathsf{D}=\left(\begin{array}{cc}\mathsf{0}&\mathsf{D}_{\rm off}\\
\mathsf{D}_{\rm off}^{\rm t}&\mathsf{D}_{\rm B}\end{array}\right)~.
\end{equation} 

$\mathsf{0}$ is a $2\times 2$ null matrix, and the other block matrices are given as follows: 
\begin{eqnarray}
\mathsf{F}_{\rm S}&=&\left(
\begin{array}{cc}
    0 & -\omega_0 \\
    \omega_0 & 0
\end{array}
\right),~\nonumber\\
\mathsf{F}_{\rm{off}}&=&\left(
\begin{array}{ccccc}
    0 & 0 &0&0&\cdots\\
  0&\gamma_1 & 0&\gamma_2 & \cdots
\end{array}
\right),~\nonumber\\
\widetilde{\mathsf{F}}_{\rm{off}}^{\rm t}&=&\left(
\begin{array}{ccccc}
 -\gamma_1 & 0&-\gamma_2& 0&\cdots\\
    0 & 0 &0&0&\cdots
\end{array}
\right),~~\nonumber\\
\mathsf{D}_{\rm{off}}&=&\frac{1}{8}\left(
\begin{array}{ccccc}
\gamma_1 & 0&\gamma_2 & 0&\cdots\\
    0&-\gamma_1 & 0 &-\gamma_2 &\cdots
\end{array}
\right),~~\label{blocks}\\
\mathsf{F}_{\rm{B}}&=&\left(
\begin{array}{ccccc}
\mu_1/2 & -\omega_1&0 & 0&\cdots\\
    \omega_1&\mu_1/2 & 0 &0 &\cdots\\
    0&0& \mu_2/2&-\omega_2&\cdots\\
    0&0&\omega_2&\mu_2/2&\cdots\\
    0&0&0&0&\mu_3/2
\end{array}
\right),~~
\label{FB}\nonumber\\
\mathsf{D}_{\rm{B}}&=&\frac{1}{4}\left(
\begin{array}{ccccc}
N_1\mu_1 & 0&0 & 0&\cdots\\
    0&N_1\mu_1 & 0 &0 &\cdots\\
    0&0& N_2\mu_2&0&\cdots\\
    0&0&0&N_2\mu_2&\cdots\\
    0&0&0&0&N_3\mu_3
\end{array}
\right)~.\nonumber
\end{eqnarray}
The block components are given
\beq
\begin{split}
\mathsf{F}_\mathrm{off}^i\!=\!\left(\!\begin{array}{cc}0&0\\0&\gamma_i\end{array}\!\right)\!,~
\tilde{\mathsf{F}}_\mathrm{off}^i\!=\!\left(\!\begin{array}{cc}-\gamma_i&0\\0&0\end{array}\!\right)\!,~
\mathsf{D}_\mathrm{off}^i\!=\!\frac{1}{8}\left(\!\begin{array}{cc}\gamma_i&0\\0&-\gamma_i\end{array}\!\right)\\
\mathsf{F}_\mathrm{B}^i\!=\!\left(\!\begin{array}{cc}\mu_i/2&-\omega_i\\\omega_i&\mu_i/2\end{array}\!\right)\!,~
\mathsf{D}_\mathrm{B}^i\!=\!\frac{1}{4}\left(\!\begin{array}{cc}N_i\mu_i&0\\0&N_i\mu_i\end{array}\!\right)~~~~~
\end{split}
\label{block_comp}
\eeq
where
$\mathsf{F}_{\rm{B}}$ and $\mathsf{D}_{\rm{B}}$ come from $\mathcal{D}_B^\mathrm{LB}$. For the RF-type dissipator, the $i$-th blocks of them are given by
\beq
\mathsf{F}_\mathrm{B}^i\!=\!\left(\!\begin{array}{cc}0&-\omega_i\\\omega_i+\alpha_i&\mu_i\end{array}\!\right)\!,~
\mathsf{D}_\mathrm{B}^i\!=\!\left(\!\begin{array}{cc}0&\delta_i\\\delta_i&N_i\mu_i/2\end{array}\!\right)~~~~~\\
\label{FB_LF}
\eeq
where 
\beq
\alpha_i=\mu_0\omega_i P(\omega_i),~\delta_i=\mu_0\omega_i Q(\omega_i)
\eeq
with
\beqn
P(\omega_i)&=&\frac{1}{\pi}\mathrm{P}\!\int_0^{\omega_c}\!\!\mathrm{d}\Omega J(\Omega)\frac{2\Omega}{\omega_i^2-\Omega^2}\\
\label{P_value}
Q(\omega_i)&=&\frac{1}{4\pi}\mathrm{P}\!\int_0^{\omega_c}\!\!\mathrm{d}\Omega J(\Omega)\left(\frac{1}{\omega_i+\Omega}+\frac{2\omega_i N(\Omega)}{\omega_i^2-\Omega^2}\right)~.
\eeqn
Note that $\mu_i=\mu_0\omega_iJ(\omega_i)$ and $\mu_0=2\pi^2M/(3V)$.

Mathematically, it describes the Ornstein-Uhlenbeck (OU) process in $N\to\infty$ dimensions. The formal solution is given~\cite{kwonPRE2011} as
\beq
\begin{split}
P(\mathbf{q},t)=\int\mathrm{d}\mathbf{q}_0P(\mathbf{q},t|\mathbf{q}_0,0)P(\mathbf{q}_0,0)~~~~~~~\\
P(\mathbf{q},t|\mathbf{q}_0,0)\propto e^{-(1/2)(\mathbf{q}-e^{-\mathsf{F} t}\mathbf{q}_0)^\mathrm{t}\cdot\mathsf{A}(t)\cdot(\mathbf{q}-e^{-\mathsf{F} t}\mathbf{q}_0)}\\
\mathsf{A}^{-1}(t)=\mathsf{A}^{-1}(\infty)-e^{-\mathsf{F}t}\mathsf{A}^{-1}(\infty)e^{-\mathsf{F}^\mathrm{t}t}\\
\mathsf{A}(\infty)=(\mathsf{D}+\mathsf{Q})^{-1}\mathsf{F}~~~~~~~~
\end{split}
\label{formal_solution}
\eeq
where the anti-symmetric matrix $\mathsf{Q}$~\cite{kwon_ao_thouless} is determined from
\beq
\mathsf{FQ+QF}^\mathrm{t}=\mathsf{FD-DF}^\mathrm{t}\label{anti_symm}~.
\eeq

The relaxation (dissipation) factor is given by $e^{-\mathsf{F}t}$ accompanied by $e^{-\mathsf{F}^\mathrm{t}t}$, so the inverse relaxation times are given by two times the real parts of the eigenvlaues of $\mathsf{F}$. There are two kind eigenvalues, $\lambda_0$ and $\{\lambda_i\}$ for $i>0$, which give the inverse relaxation times, $\tau_\mathrm{S}^{-1}$ for S and $\{\tau_\mathrm{B}^{-1}(\omega_i)\}$ for B, respectively. It is convenient to introduce dimensionless parameter $\bar{\gamma}_i=\omega_0^{-1}\gamma_i$, $\bar{\mu}=\omega_0^{-1}\mu_i$. Using the perturbation expansion of $\mathrm{det}(\mathsf{F}-\lambda \mathsf{I})=0$ for the number $N=2$ of photon modes up to $\mathcal{O}(\bar{\gamma}_i^2)$ and generalizing to $N\to\infty$, 
\beqn
\mathrm{Re}\lambda_i&\simeq&\frac{\omega_0}{2}\bar{\mu}_i +\mathcal{O}(\bar{\gamma}_i^2)\\
\mathrm{Re}\lambda_0&\simeq&\!\! \frac{\omega_0}{2}\!\!\sum_i \!\!\frac{\bar{\gamma}_i^2\bar{\mu}_i\xi_i^2}{(\xi_i^2-1)^2+\bar{\mu}_i^2(\xi_i^2+\xi_i^4)/2+\bar{\mu}_i^4\xi_i^4}\label{integral}
\eeqn
where $\sum_i$ comes from $N\to\infty$ limit and $\xi_i=\omega_i/\omega_0$. In continuum limit of $\vk$, $\sum_i\to V/(2\pi)^3\int \mathrm{d}^3 \vk$. Using $\bar{\gamma}_{\kl}=\sqrt{(\pi\nu/V)\xi} e^x_{\kl}$ from Eq.~(\ref{gamma}) for continuous variable $\xi$ for $\xi_i$. In $\bar{\mu}(\omega_k)\to 0$ limit, the integrand from Eq.~(\ref{integral}) goes to $\pi\delta(\xi^2-1)=\pi\delta(\xi-1)/2$. Given $\vk$, the angular integral of $\sum_\lambda(e^x_{\kl})^2$ is equal to $8\pi/3$. Using $\int \mathrm{d}k k^2=(\omega_0/c)^3\int \mathrm{d}\xi \xi^2$, we obtain
\beq
\tau_\mathrm{S}^{-1}=\frac{\omega_0\nu}{6}\left(\frac{\omega_0}{c}\right)^3=\frac{\omega_0\bar{\nu}}{6}
\label{tau_S}
\eeq
where $\bar{\nu}=\nu (\omega_0/c)^3$, that is dimensionless. For each photon mode,
\beq
\tau_\mathrm{B}^{-1}(\omk)=\omega_0\bar{\mu}(\omk)~.
\label{tau_B}
\eeq
In Eq.~(11) in the text, we get for typical experimental parameters
\beq
\tau_\mathrm{S}\gg\tau_\mathrm{B}\gg\omega_0^{-1}
\label{fast and slow}
\eeq
It gives rise to a desired criterion of {\it fast and slow} time scales in which bath's
variables relax much faster than system's. It is similar to a classical stochastic dynamics in phase space $(x,p)$ in overdamped limit where momentum $p$ is a fast variable compared to position $x$. 

\section{Marginalization: averaging over bath variables in $P$-representation}

We find the marginal $P$-distribution for S by averaging over B variables. It is be comparable to the derivation of classical Fokker-Planck equation by averaging over momentum variables. 
The $P$-distribution for the bare bath with no interaction with the system obeys 
\begin{equation}
\dot{P}_{\rm B}=\partial_{\mathbf{R}}\cdot\left(\mathsf{F}_{\rm{B}}
\cdot\mathbf{R}+\mathsf{D}_{\rm B}\cdot\partial_{\mathbf{R}}\right)P_{\rm B},
\end{equation}
where $\mathbf{R}=(x_1,~y_1,~x_2,~x_3,~\cdots)^{\rm t}$ and matrices are given in Eq.~(\ref{blocks}).
The steady state solution can be found to be proportional to $e^{-(1/2)\mathbf{R}^{\rm t}\cdot \mathsf{A}_{\rm B}\cdot\mathbf{R}}$.  The condition $\dot{P}_\mathrm{B}=0$
leads to 
\beq
\begin{split}
\mathrm{Tr}(\mathsf{F}_\mathrm{B}-\mathsf{D}_\mathrm{B}\mathsf{A}_\mathrm{B})=0~~~~~\\
 \mathbf{R}^\mathrm{t}\cdot\mathsf{A}_\mathrm{B}(\mathsf{F}_\mathrm{B}-\mathsf{D}_\mathrm{B}\mathsf{A}_\mathrm{B})\mathbf{R}=0~.
\label{A_B condition}
\end{split}
\eeq
$\mathsf{F}_\mathrm{B}=(\mathsf{D}_\mathrm{B}+\mathsf{Q}_\mathrm{B})\mathsf{A}_\mathrm{B}$ for antisymmetric matrix $\mathsf{Q}_\mathrm{B}$ satisfies both conditions. $\mathsf{A}_\mathrm{B}$ should be symmetric, so one can find the constraint equation for $\mathsf{Q}_\mathrm{B }$
\beq
\mathsf{F}_\mathrm{B}\mathsf{Q}_\mathrm{B}+\mathsf{Q}_\mathrm{B}\mathsf{F}_\mathrm{B}^\mathrm{t}=\mathsf{F}_\mathrm{B}\mathsf{D}_\mathrm{B}-\mathsf{D}_\mathrm{B}\mathsf{F}_\mathrm{B}^\mathrm{t}.
\eeq
which is the same kind of antisymmetric matrix as in Eq.~(\ref{anti_symm}) and can be found easily in this case. From $\mathsf{A}_\mathrm{B}=(\mathsf{D}_\mathrm{B}+\mathsf{Q}_\mathrm{B})^{-1}\mathsf{F}_\mathrm{B}$, we find in the zeroth order of $\bar{\mu}_i$
\begin{eqnarray}
\mathsf{A}_{\rm{B}}&=&\left(
\begin{array}{ccccc}
2/N_1& 0&0 & 0&\cdots\\
    0&2/N_1 & 0 &0 &\cdots\\
    0&0& 2/N_2&0&\cdots\\
    0&0&0&2/N_2&\cdots\\
    0&0&0&0&2/N_3
\end{array}
\right).
\end{eqnarray}
The higher order in $\bar{\mu}_i$ gives rise to the correction to equilibrium for B under influence of SU, which is not necessary for marginal distribution of S to be determined in the limit $\bar{\mu}\to 0$. 
In inverse $P$-representation, it corresponds to B's equilibrium density matrix $\propto e^{-\beta H_{\rm B}}$. 

As the interaction between S and B is turned on, the composite system relaxes to a steady state through a very complicated process with multiple relaxation rates. If there exist two separable \emph{fast and slow} time scales in Eq.~(\ref{fast and slow}), dynamical process becomes much simpler. $\mathbf{R}$ relaxes fast to a local equilibrium at $\mathbf{R}=\mathbf{u}$ which is slowly varying with S state $\mathbf{r}=(x,~y)^{\rm t}$. It is convenient to investigate the conditional P-distribution, because it is Gaussian as seen in Eq.~(\ref{formal_solution}). 

We conjecture 
\begin{eqnarray}
\lefteqn{P(\mathbf{r},\mathbf{R},t|\mathbf{r}_{0},0)=\frac{[{\rm det}(\mathsf{A}_{\rm B}+\mathsf{\Gamma})]^{1/2}}{(2\pi)^M}P_{\rm S}(\mathbf{r},t|\mathbf{r}_0,0) }\nonumber\\
&&~~~~~~~\times \exp\left[-\frac{1}{2}(\mathbf{R}-\mathbf{u})^{\rm t}\cdot(\mathsf{A}_{\rm B}+\mathsf{\Gamma})\cdot(\mathbf{R}-\mathbf{u})\right]\
\label{conjecture}
\end{eqnarray}
where the dependence of the initial state of B is assumed to be decayed away and only the initial state $\mathbf{r}_0$ for S appears. Note that $\mathbf{u}$ should be linear in $\mathbf{r}$ and $\mathbf{r}_0$.

Writing $\mathbf{q}=(\mathbf{r},\mathbf{R})^\mathrm{t}$, we plug the conjectured distribution in Eq.~(\ref{conjecture}) into the pseudo-FP equation in Eq.~(\ref{OU_process}) to determine $\mathbf{u}$ and $\mathsf{\Gamma}$. In this course we note that $\mathsf{F}_{\rm off}$, $\widetilde{\mathsf{F}}_{\rm off}$, $\mathsf{D}_{\rm off}$ are of $\mathcal{O}(\bar{\gamma}_i)$. $\mathbf{u}$ and $\mathsf{\Gamma}$ are born from $\HI$. We treat  $\mathbf{u}\sim\mathcal{O}(\bar{\gamma}_i)$ and $\mathsf{\Gamma}\sim\mathcal{O}(\bar{\gamma}_i^2)$, which can be seen selfconsistently. 

We expand terms resulting from Eq.~(\ref{OU_process}) as a polynomial of $\widetilde{\mathbf{R}}$. The left-hand-side leads to $\mathrm{LHS}=\dot{P}/P$ 
\beq
\mathrm{LHS}=\frac{1}{2}\mathrm{Tr}\widetilde{\mathsf{A}}_\mathrm{B}^{-1}\dot{\mathsf{\Gamma}}+\frac{\dot{P}_\mathrm{S}}{P_\mathrm{S}}
+\widetilde{\mathbf{R}}^{\rm t}\cdot\widetilde{\mathsf{A}}_\mathrm{B}\cdot\dot{\mathbf{u}}-\frac{1}{2}\widetilde{\mathbf{R}}\cdot\dot{\mathsf{\Gamma}}\cdot\widetilde{\mathbf{R}}
\eeq
where $\widetilde{\mathbf{R}}=\mathbf{R}-\mathbf{u}$ and $\widetilde{\mathsf{A}}_\mathrm{B}=\mathsf{A}_\mathrm{B}+\mathsf{\Gamma}$. $P_\mathrm{S}$ is a short notation of $P_\mathrm{S}(\mathbf{r},t|\mathbf{r}_0,0)$. The right-hand-side of Eq.~(\ref{OU_process}) is more complicated. After some algebra, we find 
$\mathrm{RHS}=[\partial_\mathbf{q}\cdot(\mathsf{F}\cdot{\mathbf{q}}+\mathsf{D}\dot\partial_\mathbf{q})P]/P$ 
\beq
\mathrm{RHS}=K_0+\widetilde{\mathbf{R}}^\mathrm{t}\cdot\mathbf{K}_1+\widetilde{\mathbf{R}}^\mathrm{t}\cdot\mathsf{K}_2\cdot\widetilde{\mathbf{R}}
\eeq
where
\beqn
K_0&=&\left[\partial_\mathbf{r}\cdot(\mathsf{F}_\mathrm{S}\cdot\mathbf{r}+\mathsf{F}_\mathrm{off}\cdot\mathbf{u})P_\mathrm{S}\right]/P_\mathrm{S}\nonumber\\
&&-\mathrm{Tr}\mathsf{D}_\mathrm{B}\mathsf{\Gamma}-\partial_\mathbf{r}\cdot(\mathsf{F}_\mathrm{off}-2\mathsf{D}_\mathrm{off}\widetilde{\mathsf{A}}_\mathrm{B})\cdot\mathbf{u}\\
\mathbf{K}_1&=&(\mathsf{F}_\mathrm{off}^\mathrm{t}-2\widetilde{\mathsf{A}}_\mathrm{B}\mathsf{D}_\mathrm{off}^\mathrm{t})\!\cdot\!\partial_\mathbf{r} \ln P_\mathrm{S}\!-\!\widetilde{\mathsf{A}}_\mathrm{B}(\widetilde{\mathsf{F}}_\mathrm{off}\cdot\mathbf{r}+\mathsf{F}_\mathrm{B}\cdot\mathbf{u})\nonumber\\
&&+\left[(\mathsf{F}_\mathrm{S}\cdot\mathbf{r}+\mathsf{F}_\mathrm{off}\cdot\mathbf{u})^\mathrm{t}\cdot\partial_\mathbf{r}\right]\widetilde{\mathsf{A}}_\mathrm{B}\cdot\mathbf{u}\\
\mathsf{K}_2&=&(\mathsf{F}_\mathrm{off}^\mathrm{t}\!-\!2\widetilde{\mathsf{A}}_\mathrm{B}\mathsf{D}_\mathrm{off}^\mathrm{t})(\partial_\mathbf{r}\mathbf{u}^\mathrm{t}) \widetilde{\mathsf{A}}_\mathrm{B}\!-\!\widetilde{\mathsf{A}}_\mathrm{B}(\mathsf{F}_\mathrm{B}\!-\!\mathsf{D}_\mathrm{B}\widetilde{\mathsf{A}}_\mathrm{B})
\eeqn
Comparing $\mathrm{LHS}$ and $\mathrm{RHS}$, we can find
\beq
\dot{\mathbf{u}}=\widetilde{\mathsf{A}}^{-1}_\mathrm{B}\cdot\mathbf{K}_1~,~~\dot{\mathsf{\Gamma}}=-(\mathsf{K}_2+\mathsf{K}_2^\mathrm{t})~.
\eeq
The differential equation for $\mathbf{u}$ is closed, but that for $\mathsf{\Gamma}$ requires $\mathbf{u}$. $\mathsf{\Gamma}$ shows how the state of B is deviated from equilibrium conversely from S. Since we are mainly interested in reduced dynamics of S, we do not investigate $\mathsf{\Gamma}$ in detail. 

Averaging over $\widetilde{\mathbf{R}}$ and using $\langle\widetilde{\mathbf{R}}\widetilde{\mathbf{R}}^\mathrm{t}\rangle=\widetilde{\mathsf{A}}_\mathrm{B}^{-1}$, we get
\beqn
\lefteqn{\dot{P}_\mathrm{S}/P_\mathrm{S}=K_0+\mathrm{Tr}\widetilde{\mathsf{A}}_\mathrm{B}^{-1}\mathsf{K}_2}\nonumber\\
&=&\left[\partial_\mathbf{r}\!\cdot\!(\mathsf{F}_\mathrm{S}\cdot\mathbf{r}+\mathsf{F}_\mathrm{off}\cdot\mathbf{u})P_\mathrm{S}\right]/P_\mathrm{S}
\!-\!\mathrm{Tr}(\mathsf{F}_\mathrm{B}\!-\!\mathsf{D}_\mathrm{B}\mathsf{A}_\mathrm{B}).
\eeqn
where the last term in the second line vanishes from Eq. (\ref{A_B condition}). Then, we find the marginal differential equation for S 
\beq
\dot{P}_\mathrm{S}(\mathbf{r},t|\mathbf{r}_0,0)=\partial_\mathbf{r}\!\cdot\!(\mathsf{F}_\mathrm{S}\cdot\mathbf{r}+\mathsf{F}_\mathrm{off}\cdot\mathbf{u})P_\mathrm{S}(\mathbf{r},t|\mathbf{r}_0,0).
\label{constrained eq}
\eeq
Up to $\mathcal{O}(\bar{\gamma}_i)$, the differential equation for $\mathbf{u}$ is given as
\beqn
\dot{\mathbf{u}}&=&(\mathsf{A}_\mathrm{B}^{-1}\mathsf{F}_\mathrm{off}^\mathrm{t}-2\mathsf{D}_\mathrm{off}^\mathrm{t})\!\cdot\!\partial_\mathbf{r} \ln P_\mathrm{S}(\mathbf{r},t|\mathbf{r}_0,0)\nonumber\\
&&-(\widetilde{\mathsf{F}}_\mathrm{off}\cdot\mathbf{r}+\mathsf{F}_\mathrm{B}\cdot\mathbf{u})+
\mathbf{r}^\mathrm{t}\cdot\mathsf{F}_\mathrm{S}\cdot\partial_\mathbf{r}^\mathsf{t}\mathbf{u}~.
\label{u_constraint}
\eeqn
Since $\mathsf{F}_\mathrm{off}\sim\mathcal{O}(\bar{\gamma}_i)$, it can determine $P_\mathrm{S}$ up to $\mathcal{O}(\bar{\gamma}_i^2)$, corresponding to the second order of $\HI$. 

\section{Derivation of the unconstrained marginal differential equation }  

We have derived the constrained marginal differential equation (\ref{constrained eq}) with the constraint in Eq.~(\ref{u_constraint}) for $\mathbf{u}$. It is non-trivial to solve the two equations simultaneously. We further conjecture the conditional P-distribution for S as 
\begin{eqnarray}
\lefteqn{P_{\rm S}(\mathbf{r},t|\mathbf{r}_0,0)=\frac{[{\rm{det}\mathsf{A}}_\mathrm{S}(t)]^{1/2}}{2\pi}}\nonumber\\
&&\times \exp\!\!\left[-\frac{1}{2}(\mathbf{r}\!-\!e^{-\mathsf{F}_{\rm eff} t}\!\cdot\mathbf{r}_0)^{\rm t}\!\!\cdot\!\mathsf{A}_\mathrm{S}(t)\!\cdot\!(\mathbf{r}\!-\!e^{-\mathsf{F}_{\rm eff} t}\!\cdot\mathbf{r}_0\!)\right]~~~
\label{conditional_PS}
\end{eqnarray}
which is motivated from the classical FP equation. The initial condition $\mathsf{A}^{-1}_\mathrm{S}(0)=\mathsf{0}$ is required to get $P_{\rm S}(\mathbf{r},0|\mathbf{r}_0,0)=\delta(\mathbf{r}-
\mathbf{r}_0)$. Since $\mathbf{u}$ is linear in $\mathbf{r}$ and $\mathbf{r}_0$, we write 
\begin{equation}
\mathbf{u}=\mathsf{A}_{\rm off}\cdot\mathbf{r}+\mathsf{B}_{\rm off}e^{-\mathsf{F}_{\rm eff} t}\cdot\mathbf{r}_0
\label{u_form}
\end{equation}
where $\mathsf{A}_{\rm off},~\mathsf{B}_{\rm off}$ are $2N\times 2$ column matrices for the number $N$ for photon modes. In continuum limit of $\kl$, $N$ goes to $\infty$. 

We plug Eqs.~(\ref{conditional_PS}) and (\ref{u_form}) into Eqs.~(\ref{constrained eq}) and (\ref{u_constraint}). First, we get
\begin{eqnarray}
 {\rm Tr}\dot{\mathsf{A}}_\mathrm{S}\mathsf{A}_\mathrm{S}^{-1}&=&2{\rm Tr}(\mathsf{F}_{\rm S}+\mathsf{F}_{\rm off}\mathsf{A}_{\rm off})~,\\
\mathsf{F}_{\rm eff}&=&\mathsf{F}_{\rm S}+\mathsf{F}_{\rm off}(\mathsf{A}_{\rm off}+\mathsf{B}_{\rm off})~,
\label{F_effective}\\
\dot{\mathsf{A}}_\mathrm{S}&=&\mathsf{A}_\mathrm{S}(\mathsf{F}_{\rm S}+\mathsf{F}_{\rm off}\mathsf{A}_{\rm off})\!-\!(\mathsf{F}_{\rm S}\!-\!\mathsf{A}_{\rm off}^{\rm t}\mathsf{F}_{\rm off}^{\rm t})\mathsf{A}_\mathrm{S}~.~~.
\label{AS_rate}
\end{eqnarray}
Noting $\mathsf{F}_{\rm S}^{\rm t}=-\mathsf{F}_{\rm S}$, the first and last equations are consistent. 

Next, we get
\begin{eqnarray}
\dot{\mathsf{A}}_{\rm off}&=&-\mathsf{F}_{\rm B}\mathsf{A}_{\rm off}+\mathsf{A}_{\rm off}\mathsf{F}_{\rm S}-\widetilde{\mathsf{F}}_{\rm off}-\widetilde{\mathsf{D}}\mathsf{A}_\mathrm{S}(t)
\label{Aoff}\\
\dot{\mathsf{B}}_{\rm off}&=&-\mathsf{F}_{\rm B}\mathsf{B}_{\rm off}+\mathsf{B}_{\rm off}\mathsf{F}_{\rm eff}+\widetilde{\mathsf{D}}\mathsf{A}_\mathrm{S}(t)
\label{Boff}
\end{eqnarray}
where $\widetilde{\mathsf{D}}=\mathsf{A}_{\rm B}^{-1}\mathsf{F}_{\rm off}^{\rm t}-2\mathsf{D}_{\rm off}^{\rm t}$. From Eq.~(\ref{F_effective}), $\mathsf{F}_{\rm eff}$ in Eq.~(\ref{Boff}) is approximately equal to $\mathsf{F}_{\rm S}$. Then, we find
\beq
\dot{\mathsf{A}}_{\rm off}+\dot{\mathsf{B}}_{\rm off}=-\mathsf{F}_{\rm B} (\mathsf{A}_{\rm off}+\mathsf{B}_{\rm off})+(\mathsf{A}_{\rm off}+\mathsf{B}_{\rm off})\mathsf{F}_{\rm S}-\widetilde{\mathsf{F}}_{\rm off}
\label{Aoff+Boff}
\eeq

We consider S and B to be initially in a product state with interaction not yet turned on. Then, $\mathsf{A}_{\rm off}$ and $\mathsf{B}_{\rm off}$ are equal to zero at $t=0$.  Upon this initial condition, Eqs.~ (\ref{Boff}) and (\ref{Aoff+Boff}) yield
\begin{eqnarray}
\mathsf{A}_{\rm off}+\mathsf{B}_{\rm off}&=&-\int_{0}^t dt' e^{-\mathsf{F}_{\rm B}t'}\widetilde{\mathsf{F}}_{\rm off} e^{\mathsf{F}_{\rm S}t'}=-\mathsf{I}_1~,
\label{ABoff_integral}\\
\mathsf{B}_{\rm off}&=&\int_{0}^t dt' e^{-\mathsf{F}_{\rm B}t'}\widetilde{\mathsf{D}} \mathsf{A}(t-t')e^{\mathsf{F}_{\rm S}t'}~.
\label{Boff_integral}
\end{eqnarray}
As noticed previously,  $e^{-\mathsf{F}_{\rm B}t'}$ decays very fast as $e^{-t/\tau_B(\omega_i)}$ in long relaxation time scale of S $\sim\tau_S$. In Eq.~(\ref{Boff_integral}), $\mathsf{A}(t-t')$ is needed only for small $t'\ll \tau_S$. In this short time range, Eq.~(\ref{AS_rate}) gives 
\beq
\dot{\mathsf{A}}_\mathrm{S}=\mathsf{A}_\mathrm{S}\mathsf{F}_{\rm S}-\mathsf{F}_{\rm S}\mathsf{A}_\mathrm{S}+\mathcal{O}(\bar{\gamma}_i^2)~.
\eeq
Therefore, we have
\begin{equation}
\mathsf{A}_\mathrm{S}(t-t')\simeq e^{\mathsf{F}_{\rm S}t'}\mathsf{A}_\mathrm{S}(t)e^{-\mathsf{F}_{\rm S}t'}~,
\end{equation}
which is a behavior from pure quantum oscillation. Then, Eq.~(\ref{Boff_integral}) becomes
\begin{equation}
\mathsf{B}_{\rm off}=\int_{0}^t dt' e^{-\mathsf{F}_{\rm B}t'}\widetilde{\mathsf{D}}e^{\mathsf{F}_{\rm S} t'} \mathsf{A}_\mathrm{S}(t)=\mathsf{I}_2 \mathsf{A}_\mathrm{S}(t) ~.
\label{Boff_I2}
\end{equation} 

The two integrals, $\mathsf{I}_1$ and $\mathsf{I}_2$, can be rewritten by integrating by parts. We get 
\beqn
\mathsf{I}_1&=&-\mathsf{F}_\mathrm{B}^{-1}\left[e^{-\mathsf{F}_{\rm B}t'}\widetilde{\mathsf{F}}_{\rm off} e^{\mathsf{F}_{\rm S}t'}\Big|_0^t-\mathsf{I}_1\mathsf{F}_\mathrm{S}\right]~,
\\
\label{ineg_by_part1}
\mathsf{I}_2&=&-\mathsf{F}_\mathrm{B}^{-1}\left[e^{-\mathsf{F}_{\rm B}t'}\widetilde{\mathsf{D}} e^{\mathsf{F}_{\rm S}t'}\Big|_0^t-\mathsf{I}_1\mathsf{F}_\mathrm{S}\right]~.
\label{integ_by_part2}
\eeqn
Neglecting $e^{-\mathsf{F}_\mathrm{B}t}$, we have 
\begin{eqnarray}
\mathsf{F}_{\rm B}\mathsf{I}_1-\mathsf{I}_1\mathsf{F}_{\rm S}&=&\widetilde{\mathsf{F}}_{\rm off}~,\\
\mathsf{F}_{\rm B}\mathsf{I}_2-\mathsf{I}_2\mathsf{F}_{\rm S}&=&\widetilde{\mathsf{D}}~.
\end{eqnarray}
In terms of block matrices, 
\begin{eqnarray}
\mathsf{F}_{\rm B}^i\mathsf{I}_1^i-\mathsf{I}_1^i\mathsf{F}_{\rm S}&=&\widetilde{\mathsf{F}}_{\rm off}^i~,
\label{I1}\\
\mathsf{F}_{\rm B}^i\mathsf{I}_2^i-\mathsf{I}_2^i\mathsf{F}_{\rm S}&=&\left((\mathsf{A}_\mathrm{B}^i)^{-1}(\mathsf{F}_\mathrm{off}^i)^\mathrm{t}-2(\mathsf{D}_\mathrm{off}^i)^\mathrm{t}\right)~,
\label{I2}
\end{eqnarray}
where block matrices for $i$-th mode are given in Eqs.~(\ref{block_comp}) and (\ref{FB_LF}). Especially, we recall 
$\mathsf{F}_\mathrm{B}^i$ depending on the type of thermostat
\beq
\mathsf{F}_\mathrm{B}^i=\left(\begin{array}{cc}
\epsilon_3\mu_i&-\omega_i\\
\omega_i+\epsilon_2\alpha_i &\epsilon_1\mu_i
\end{array}\right)
\eeq
where $\mu_i=\mu_0\omega_iJ(\omega_i)$ and $\alpha_i=\mu_0\omega_i P(\omega_i)$ from Eq.~(\ref{P_value}).
$\epsilon_1=\epsilon_3=1/2,~\epsilon_2=0$ for the LB-type and $\epsilon_1=\epsilon_2=1,~\epsilon_3=0$ for the RF-type.
Then, elements of $2\times 2$ matrices $\mathsf{I}_1^i,~\mathsf{I}_2^i$ can be found by solving linear equations in Eqs.~(\ref{I1}) and (\ref{I2}), which determines $\mathbf{u}$ in Eq.~(\ref{u_form}).

We obtain the unconstrained marginal differential equation by plugging the solved $\mathbf{u}$ into Eq.~(\ref{constrained eq}) 
\begin{eqnarray}
\lefteqn{\dot{P_S}(\mathbf{r},t|\mathbf{r}_0,0)=\partial_{\mathbf{r}}\cdot\Big[\left\{\mathsf{F}_{\rm S}+\mathsf{F}_{\rm off}(\mathsf{A}_{\rm off}+\mathsf{B}_{\rm off})\right\}\cdot\mathbf{r}}
\nonumber\\
&&-\mathsf{F}_{\rm off}\mathsf{B}_{\rm off}\cdot(\mathbf{r}-e^{-\mathsf{F}_\mathrm{eff}}\cdot\mathbf{r}_0)\Big]P_S(\mathbf{r},t|\mathbf{r}_0,0)\nonumber\\
&=&\partial_{\mathbf{r}}\cdot\left(\mathsf{F}_{\rm eff}\cdot\mathbf{r}+\mathsf{F}_\mathrm{off}\mathsf{I}_2\cdot\partial_{\mathbf{r}}\right)P_S(\mathbf{r},t|\mathbf{r}_0,0)~.
\label{final_master}
\end{eqnarray}
where Eqs.~(\ref{F_effective}) and (\ref{Boff_I2}) are used. Note that the antisymmetric part of $\mathsf{F}_\mathrm{off}\mathsf{I}_2$ vanishes. 
Finally, we write
\beq
\dot{P}_\mathrm{S}(\mathbf{r},t)=\partial_\mathbf{r}\cdot\left(\mathsf{F}_{\rm eff}\cdot\mathbf{r}+\mathsf{D}_\mathrm{eff}\cdot\partial_{\mathbf{r}}\right)P_S(\mathbf{r},t)
\eeq
where effective drift and diffusion matrices are given as
\begin{eqnarray}
\mathsf{F}_{\rm eff}&=&\mathsf{F}_{\rm S}-\mathsf{F}_{\rm off}\mathsf{I}_1~,\label{Feff}\\
\mathsf{D}_{\rm eff}&=&\frac{\mathsf{F}_{\rm off} \mathsf{I}_2+\mathsf{I}_2^{\rm t}\mathsf{F}_{\rm off}^{\rm t}}{2}~.\label{Deff}
\end{eqnarray} 
Using block matrices, we have
\begin{eqnarray}
\mathsf{F}_{\rm eff}&=&\mathsf{F}_{\rm S}-\sum_i\mathsf{F}_{\rm off}^i\mathsf{I}_1^i~,\label{Feff_block}\\
\mathsf{D}_{\rm eff}&=&\sum_i\frac{\mathsf{F}_{\rm off}^i \mathsf{I}_2^i+(\mathsf{I}_2^i)^{\rm t}(\mathsf{F}_{\rm off}^{i})^{\rm t}}{2}~.\label{Deff_block}
\end{eqnarray} 
Then, we have $P_\mathrm{S}(\mathbf{r},t)=\int\mathrm{d}\mathbf{r}_0 P_S(\mathbf{r},t|\mathbf{r}_0,0)P_\mathsf{S}(\mathbf{r}_0,0)$.

\section{The solution of the marginal differential equation}

From Eqs.~(\ref{I1}) and (\ref{I2}), $\mathsf{I}_{1}$ and $\mathsf{I}_{2}$ are $\mathcal{O}(\bar{\gamma}_i)$. So, the matrix elements of $\mathsf{F}_{\rm eff}$ and $\mathsf{D}_{\rm eff}$ are sums over terms proportional to $\bar{\gamma}_i^2$, as observed in Eqs.~(\ref{Feff_block}) and (\ref{Deff_block}). Then, the matrix elements of the two matrices are written as
$\sum_i \bar{\gamma}_i^2  g_i(\omega_i/\omega_0)$. In the continuum limit, 
$\sum_i$ goes to $V/(2\pi)^3\int d^3\mathbf{k}\sum_\lambda$, integrals in $\mathbf{k}$-vector space. The polarization vectors are chosen to form orthonormal basis vectors with completeness relation: $\mathbf{e}_{\mathbf{k}1}^{\rm t}\mathbf{e}_{\mathbf{k}1}+\mathbf{e}_{\mathbf{k}2}^{\rm t}\mathbf{e}_{\mathbf{k}2}+\mathbf{k}^{\rm t}\mathbf{k}/k^2=\mathsf{I}$. Given $\mathbf{k}$, $(\mathbf{e}_{\mathbf{k}1}^x)^2+(\mathbf{e}_{\mathbf{k}2}^x)^2=1-(k_x/k)^2=1- \sin^2\theta$ in spherical coordinates $(k,\theta,\phi)$, which leads the angular integral of $\sum_\lambda(e_{\kl}^x)^2$ from $\bar{\gamma}_i^2$ over $\theta$ and $\phi$ to equal to $8\pi/3$. Then, we have a basic formula of integrals:
\beqn
\sum_i \bar{\gamma}_i^2  g_i(\omega_i/\omega_0)&=&\frac{\pi\nu}{V}\sum_i (e^{x}_i)^2 (\omega_i/\omega_0) g(\omega_i/\omega_0)\nonumber\\
&\to& \frac{\bar{\nu}}{3\pi}\int_0^{\xi_c}\!\!\! d\xi~\xi^3 g(\xi)~,
\eeqn
where $k=\omega/c$, $\xi=\omega/\omega_0$, $\bar{\nu}=\nu(\omega_0/c)^3$ are used. $\xi_c=\omega_c/\omega_0$ is the upper limit  corresponding to the cut-off frequencies of the CL oscillators. 

Then, $\mathsf{F}_{\rm eff}$ and $\mathsf{D}_{\rm eff}$ are found as
\begin{equation}
\mathsf{F}_{\rm eff}=\omega_0\left(\begin{array}{cc}0&-1\\ 1+\bar{\nu} f_2 & \bar{\nu}f_1\end{array}\right)~,~\mathsf{D}_{\rm eff}=\omega_0\left(\begin{array}{cc}0& \bar{\nu}d_2\\ \bar{\nu}d_2& \bar{\nu}d_1\end{array}\right)~.
\label{simple_matrix}
\end{equation}  
We find matrix elements
\beqn
 f_1&=&\frac{1}{3\pi}\!\!\!\int_0^{\xi_c}\!\!\! \!\!d\xi ~\frac{\bar{\mu}(\epsilon_1+\epsilon_3)\xi^4+\mathcal{O}(\bar{\mu}^2)}{h(\xi)},\nonumber\\
 f_2&=&-\frac{1}{3\pi}\!\!\!\int_0^{\xi_c}\!\!\! \!\!d\xi ~\frac{\xi^4(\xi^2-1)+\mathcal{O}(\bar{\mu})}{h(\xi)},\nonumber\\
d_1&=&\frac{1}{24\pi}\!\!\!\int_0^{\xi_c}\!\!\! \!\!d\xi ~\frac{\bar{\mu}\xi^3[(1\!-\!\xi)(\epsilon_1\!-\!\xi\epsilon_3)\!+\!
   2 N(\xi) (\epsilon_1\!+\!  \epsilon_3\xi^2 )]\!+\!\mathcal{O}(\bar{\mu}^2)}{h(\xi)},\nonumber\\
d_2&=&\frac{1}{24\pi}\!\!\!\int_0^{\xi_c}\!\!\! \!\!d\xi ~\left[\frac{\xi^3(\xi^2-1)[ \xi-1 - 
   2 N(\xi)]+\mathcal{O}(\bar{\mu})}{h(\xi)}\right],
 \eeqn
where $N(\xi)=(e^{\beta\hbar\omega_0\xi}-1)^{-1}$ and 
\beqn
h(\xi)&=&(\xi^2-1)^2-2\epsilon_2\bar{\alpha}\xi(1-\xi^2) \nonumber\\
&&+\epsilon_2^2\xi^2\bar{\alpha}^2+(\epsilon_1^2+\epsilon_3^2+2\epsilon_1\epsilon_3\xi^2)\bar{\mu}^2
\eeqn
$\bar{\mu}=\mu_0\xi J(\omega_0\xi)$ and $\bar{\alpha}=\mu_0\xi P(\omega_0\xi)$ are defined in Eq.~(\ref{FB_LF}).

Now, we will take the $\bar{\mu},~\bar{\alpha}\to 0$ limit. For diagonal elements,$f_1$ and $d_1$, the integrand has a delta function in this limit, 
\beq
\frac{\sqrt{\epsilon_2^2\bar{\alpha}^2+(\epsilon_1+\epsilon_3)^2\bar{\mu}^2}}{(\xi^2 -1)^2+\epsilon_2^2\bar{\alpha}^2+(\epsilon_1+\epsilon_3)^2\bar{\mu}^2}\to \pi\delta(\xi^2-1)=\pi\delta(\xi-1)/2~.
\eeq 
For off-diagonal elements, $f_2,~d_2$ lead to the principal values by neglecting $\bar{\alpha},~\bar{\mu}$. 
For $\xi_c>1$, i.e., $\omega_c>\omega_0$, we have
\begin{eqnarray}
 f_1&=&R_\mathrm{B}\frac{\bar{\nu}}{6}~,\\
 f_2&=&-\frac{\bar{\nu}}{3\pi}~ \mathrm{P}\int_0^{\xi_c}\!\!\! \!\!d\xi ~\frac{\xi^4}{\xi^2 -1}~,\\
d_1&=&R_\mathrm{B}\frac{\bar{\nu}}{12}N_0~,\\
d_2&=&\frac{\bar{\nu}}{24\pi}~\mathrm{P}\int_0^{\xi_c}\!\!\! \!\!d\xi ~\xi^3\left[\frac{1}{ \xi+1}-\frac{2N(\xi)}{\xi^2-1} \right]~,
\label{f and d}
\end{eqnarray}
where $N_0$ is the average number of quanta for the system and 
\beq
R_\mathrm{B}=\frac{(\epsilon_1+\epsilon_3)J_0}{\sqrt{\epsilon_2^2P_0^2+(\epsilon_1+\epsilon_3)^2J_0^2}}
\eeq
with $J_0=J(\omega_0)$ and $P(\omega_0)$ in Eq.~(\ref{FB_LF}). We have
\beq
R_\mathrm{B}=\left\{\begin{array}{cc} 1~~~;&\textrm{LB-type thermostat}\\
\frac{J_0}{\sqrt{P_0^2+J_0^2}};&\textrm{LF-type thermostat}
\end{array}\right.
\eeq

Using $\mathsf{A}_\mathrm{S}^{-1}\dot{\mathsf{A}}_\mathrm{S}\mathsf{A}_\mathrm{S}^{-1}=-\dot{\mathsf{A}}_\mathrm{S}^{-1}$,  Eq.~(\ref{AS_rate}) can be rewritten as
\begin{equation}
\dot{\mathsf{A}}_\mathrm{S}^{-1}=-\mathsf{F}_{\rm eff}\mathsf{A}_\mathrm{S}^{-1}-\mathsf{A}\mathrm{S}^{-1}\mathsf{F}_{\rm eff}^{\rm t}
+2\mathsf{D}_{\rm eff}~.
\label{inverse_A}
\end{equation}
Then, we find
\begin{equation}
\mathsf{A}_\mathrm{S}^{-1}(t)=\mathsf{A}_\mathrm{S}^{-1}(\infty)-e^{-\mathsf{F}_{\rm eff} t}\mathsf{A}_\mathrm{S}^{-1}(\infty)e^{-\mathsf{F}_{\rm eff}^{\rm t} t}
\end{equation}
where the steady state kernel $\mathsf{A}_\mathrm{S}^{-1}(\infty)$ is determined from 
\begin{equation}
\mathsf{F}_{\rm eff}\mathsf{A}_\mathrm{S}^{-1}(\infty)+\mathsf{A}_\mathrm{S}^{-1}(\infty)\mathsf{F}_{\rm eff}^{\rm t}=2\mathsf{D}_{\rm eff}~. 
\end{equation}
Using the formalism in our previous study~\cite{kwonPRE2011},
\beq
\mathsf{A}_\mathrm{S}(\infty)=(\mathsf{D}_\mathrm{eff}+\mathsf{Q}_\mathrm{eff})^{-1}\mathsf{F}_\mathrm{eff}
\label{steady state}
\eeq
where the anti-symmetric matrix $\mathsf{Q}_\mathrm{eff}$ can be found from
\beq
\mathsf{F}_\mathrm{eff}\mathsf{Q}_\mathrm{eff}+\mathsf{Q}_\mathrm{eff}\mathsf{F}_\mathrm{eff}^\mathrm{t}=\mathsf{F}_\mathrm{eff}\mathsf{D}_\mathrm{eff}-\mathsf{D}_\mathrm{eff}\mathsf{Q}_\mathrm{eff}^\mathrm{t}
\eeq

\section{Comparison with the Redfield equation}
We consider a unitary dynamics for a {\it closed} composite system of S and B. The reduced density matrix for S satisfies in interaction picture
\beq
\dot{\tilde{\rho}}_\mathrm{S}(t)=-\frac{1}{\hbar^2}\mathrm{Tr}_\mathrm{B} \int_0^t\mathrm{d}t' \left[\tilde{H}_\mathrm{I}(t),\left[\tilde{H}_\mathrm{I}(t'), \tilde{\rho}_\mathrm{SB}(t')\right]\right]~.
\label{previous}
\eeq
After the usual weak coupling scheme via the Born-Markov approximation, we can derive the Redfield quantum master equation  
$\dot{\tilde{\rho}}_\mathrm{S}=\tilde{\mathcal{D}}_\mathrm{S}^\mathrm{RF}[\tilde{\rho}_\mathrm{S}]$, given in the Schr\"{o}dinger picture as
\beq
\dot{\hrho}_\mathrm{S}=-\frac{i}{\hbar}\left[\HS,\hrho_\mathrm{S}\right]+\mathcal{D}_\mathrm{S}^\mathrm{RF}[\hrho_\mathrm{S}]~.
\eeq
Here, $\mathcal{D}_\mathrm{S}^\mathrm{RF}[\hrho_\mathrm{S}]$ is a dissipator (superoperator) of S acting on $\hrho_\mathrm{S}$. It is different by definition from $\mathcal{D}_\mathrm{B}^\mathrm{RF}[\hrho_\mathrm{S}]$ derived in Sec. II. However, the derivation steps are the same as shown in the section. 

We consider $\HI$ from Eq.~(\ref{H_I}), that represents a dipole interaction between a harmonic S and photonic B,
\beq
\HI=-\sum_{\kl} \sqrt{\frac{\pi \hbar^2 q^2 \omk }{m\omega_0 V}}e_{\kl}^x
(a+a^\dagger)(b_{\mathbf{k}\lambda}-b_{\mathbf{k}\lambda}^\dagger)
\eeq
where $\omega_vk=c|\vk|$. Eigenoperators are defined from $\hat{A}=\ha+\had$ as
\beq
\hat{A}(\omega)=\!\!\!\!\!\!\!\!\!\!\!\!\!\sum_{n,n';n'\!-\!n=\omega/\omega_0 }\!\!\!\!\!\!\!\!\!\!\!\!|n\rangle\langle  n|\hat{A}| n'\rangle\langle n'| =\delta_{\omega,\omega_0} \ha
+\delta_{\omega,-\omega_0}\had~,
\eeq
with $\hat{A}^\dagger(\omega)=\hat{A}(-\omega)$. Then, we find
\beqn
\label{dissipator_RF_previous}
\mathcal{D}_\mathrm{S}^\mathrm{RF}[\hrho_\mathrm{S}]&=&\!\!\!\!\!\!\!\!\!\sum_{\omega=\pm\omega_0,\omega'=\pm\omega_0}\!\!\!\!\!\!\!\!\!\widetilde{\Gamma}(\omega)\left(\hat{A}(\omega)\hrho_\mathrm{S}\hat{A}^\dagger(\omega')\right.\nonumber\\
&&\left.-\frac{1}{2}\{ \hat{A}^\dagger(\omega')\hat{A}(\omega),\hrho_\mathrm{S}\}\right)
+\mathrm{h.c.}
\eeqn
where 
\beqn
\lefteqn{\widetilde{\Gamma}(\omega)=\sum_\kl\sum_\klp\frac{\pi q^2 \sqrt{\omk\omega_{\vk'}} }{m\omega_0 V}e_{\kl}^xe_{\klp}^x}
\nonumber\\
&&\times\!\!\int_0^\infty\!\!\!\!\! \mathrm{d}s e^{i\omega s}\!\left\langle\!(\hat{b}_{\kl}e^{-i\omk s}\!\!-\!\! \hat{b}_{\kl}^\dagger e^{i\omk s})(\hat{b}_{\klp} \!\!-\!\!\hat{b}_{\klp}^\dagger)\!\right\rangle\nonumber\\
&=&\frac{\omega_0\nu}{8\pi^2}\int\!\!\mathrm{d}^3\vk~\omega_{\vk}\!\! \sum_\lambda(e^x_{\klp})^2\Big[\pi(N(\omk\! )\!+\!1)\delta(\omega\!-\!\omk)\nonumber\\
&&+i\mathrm{P}\left(\frac{1}{\omega-\omk}+\frac{1}{\omega+\omk}\right)\Big]
\eeqn
where $\nu=q^2/(m\omega_0^2)$ and $\sum_{\vk}\to V/(2\pi)^3\int \mathrm{d}^3\vk$ for large $V$ is used. Using $\int_0^{2\pi}\mathrm{d}\phi\int_0^\pi \mathrm{d}\theta \sin\theta \sum_\lambda(e^x_\lambda)^2=8\pi/3$, we find
\beqn
\widetilde{\Gamma}(\omega)
\label{tilde_Gamma}
&=&\frac{\omega_0\bar{\nu}}{3}\Big[(N(\omega)+1)\mathrm{sign}(\omega)\nonumber\\
&&+\frac{i}{\pi}\mathrm{P}\int\frac{\mathrm{d}\omk}{\omega_0^3}~\omk^3\left(\frac{N(\omk)+1}{\omega-\omk}+\frac{N(\omega_k)}{\omega+\omk}\right)
\nonumber\\
&=&\frac{\gamma(\omega)}{2}+iS(\omega)~.
\eeqn
where $\bar{\nu}=\nu(\omega_0/c)^3$ is the same as defined in the text. 

$\omega$ is equal to either $\omega_0$ or $-\omega_0$, then $\gamma(\omega_0)=(2/3)\omega_0\bar{\nu} (N(\omega_0)+1)$ and $\gamma(-\omega_0)=(2/3)\omega_0\bar{\nu} N(\omega_0)$. Plugging $\tilde{\Gamma}(\omega)$ in Eq.~(\ref{tilde_Gamma})) into Eq.~(\ref{dissipator_RF_previous}), we find the Redfield equation 
\beq
\dot{\hrho}_\mathrm{S}=-i[\omega_0\had\ha+\hat{H}_\mathrm{shift},\hrho_\mathrm{S}]+\mathcal{D}_S^\mathrm{LB}[\hrho_\mathrm{S}]+\mathcal{D}_S^\mathrm{add}[\hrho_\mathrm{S}]
\eeq
where
\beqn
\mathcal{D}_\mathrm{S}^\mathrm{LB}[\hrho_{\rm S}]&=&\frac{2\omega_0\bar{\nu}}{3}
\!\Bigg[(N_0+1)\!\!\left( \!\hat{a}\hrho_{\rm S}\had\! \!-\frac{1}{2}\left\{\had \ha , \hrho_{\rm S}\right\} \!\!\right)\nonumber\\
&&+N_0\!\left(\! \had\hrho_{\rm S}\ha\! -\!\frac{1}{2}\!\left\{\ha\had , \hrho_{\rm S}\right\} \!\!\right)\Bigg]
\eeqn
and
\beq
\hat{H}_\mathrm{shift}=\frac{\omega_0\bar{\nu}}{3\pi}\tilde{S}_+\left(\had\ha+\frac{1}{2}(\ha\ha+\had\had)\right)
\eeq
and
\beqn
\lefteqn{\mathcal{D}_\mathrm{S}^\mathrm{add}[\hrho_{\rm S}]=\frac{\omega_0\bar{\nu}}{3}\times}\nonumber\\
&&\Bigg[(N_0+1)\Big( \ha \hrho_{\rm S} \ha+\had \hrho_{\rm S}\had-\ha\ha\hrho_{\rm S}-\hrho_{\rm S}\had \had\Big) \nonumber\\
&&+N_0\Big( \ha\hrho_{\rm S} \ha+\had \hrho_{\rm S}\had
-\had\had\hrho_{\rm S}-\hrho_{\rm S}\ha \ha\Big)\nonumber\\
&&+\frac{i}{\pi}S_{-}\Big( \ha\hrho_{\rm S} \ha
-\had \hrho_{\rm SB}\had
-\frac{1}{2}\left\{\ha\ha-\had \had ,\hat{\rho}_{\rm S}\right\}\!\Big)\!\Bigg]~~~
\eeqn
where 
\beqn
\tilde{S}_{+}&=&\mathrm{P}\!\int_0^{\xi_c}\!\!\!\!\!\mathrm{d}\xi\frac{2\xi^4}{1-\xi^2}\\
S_{-}(\omega)&=&\mathrm{P}\!\int_0^{\xi_c}\!\!\!\!\!\mathrm{d}\xi\frac{2\xi^3(2N(\xi)+1)}{1-\xi^2}~.
\eeqn
where we change a variable as $\omk/\omega_0=\xi$. We define $N_0=(e^{\beta\hbar\omega_0}-1)^{-1}$ and $N(\xi)=(e^{\beta\hbar\omega_0\xi}-1)^{-1}$

Then, we find the RF equation in the Schr\"{o}dinger picture
\beq
\dot{\hrho}_\mathrm{S}=-\frac{i}{\hbar}\left[\HS,\hrho_\mathrm{S}\right]+\mathcal{D}_\mathrm{S}^\mathrm{RF}[\hrho_\mathrm{S}]~.
\eeq
Applying the P-representation for the coherent state of $\ha$, we find
\beq
\dot{P}_S=\partial_\mathbf{r}\cdot\left(\mathsf{F}_\mathrm{RF}\cdot\mathbf{r}+\mathsf{D}_\mathrm{RF}\cdot\partial_\mathbf{r}\right)P_\mathrm{S}
\eeq
where
\begin{equation}
\mathsf{F}_{\rm RF}\!=\!\omega_0\!\left(\!\begin{array}{cc}0&-1\\ 1+\bar{\nu} f_2^\mathrm{RF} & \bar{\nu}f_1^\mathrm{RF}\end{array}\!\right),~\mathsf{D}_{\rm RF}\!=\!\left(\!\begin{array}{cc}0& \bar{\nu}d_2^\mathrm{RF}\\ \bar{\nu}d_2^\mathrm{RF}& \bar{\nu}d_1^\mathrm{RF}\end{array}\!\right)~.
\label{RF_matrix}
\end{equation}  
It is remarkable that it has the same form as our marginal equation in Eq.~(\ref{simple_matrix}). We find a simple relation of the matrix elements with ours
\beq
f_1=R_\mathrm{B}\frac{f_1^\mathrm{RF}}{4},~f_2=\frac{f_2^\mathrm{RF}}{4},~d_1=R_\mathrm{B}\frac{d_1^\mathrm{RF}}{4},~~d_2=\frac{d_2^\mathrm{RF}}{4}~.
\eeq
For example, all matrix elements from the RF master equation are 4 times larger than ours fr0m the LB-type thermostat. It is clear that the steady state of our theory is different from that of the conventional RF equation which is supposed to be the mean-force-Gibbs (MFG) state $\propto \mathrm{Tr}_\mathrm{B}e^{-\beta(\HS+\HB+\HI)}$. Difference lies in the perspectives of the two theories: S is coupled with B which exclusively coupled with SU in our theory while S and B are closed from environment in the previous theory.
Many practical situations are more relevant to our theory.

\section{Steady state }
From Eq.~(\ref{steady state}), we find
\begin{eqnarray}
\lefteqn{\mathsf{A}_\mathrm{S}(\infty)=\left(\begin{array}{cc}
\frac{f_1}{d_1}+\bar{\nu}\frac{f_1}{d_1}\frac{f_2d_1-2f_1 d_2}{d_1}&0\\0&\frac{f_1}{d_1}\end{array}\right)}\nonumber\\
&=&\left(\begin{array}{cc}
\frac{2}{N_0}+\frac{2\bar{\nu}}{N_0}\left(f_2-\frac{4}{N_0}d_2\right)&0\\0&\frac{2}{N_0}\end{array}\right) \textrm{for $\omega_c\!>\!\omega_0$}.
\end{eqnarray}
We use $f_1/d_1=2/N_0$, which is the condition for S to be in equilibrium in leading order and is satisfied independent of the type of thermostat on B. 
The deviation from equilibrium comes from higher order terms which is given in the second order in $\HI$ as
$2\bar{\nu}/N_0\left(f_2-4d_2/N_0\right)$, which is also independent of the type of thermostat. It is interesting that the steady state is independent of the thermostat originated from SU, but the time-dependent state depends on the thermostat, i.e. on how B is in thermal contact with SU. 
 
The steady state density matrix for S can be found via inverse $P$-representation from $P^\mathrm{ss}_\mathrm{S}\propto e^{-1/2 \mathbf{r}^\mathrm{t}\cdot \mathsf{A}_\mathrm{S}(\infty)\cdot\mathbf{r}}$. In polar coordinates, 
\beq
P^\mathrm{ss}_\mathrm{S}\!\!=\!\!\frac{\sqrt{(2/N_0)^2(1+c\bar{\nu})}}{2\pi}\exp\left[-\frac{r^2}{N_0}(1+C\bar{\nu}\cos^2\theta)\right]~,
\eeq
where
\beq
C=f_2-\frac{4}{N_0}d_2~.
\eeq
Then, we find the steady state density matrix for S up to $\mathcal{O}(\bar{\nu})$
\beqn
\hrhoS&=&\int \mathrm{d}^2 z|z\rangle\langle z| P^\mathrm{ss}_\mathrm{S}\nonumber\\
&=&\frac{2}{N_0}\int_0^{2\pi}\!\!\frac{\mathrm{d}\theta}{2\pi}\int_0^\infty\!\!\!\mathrm{d}r e^{-\left(1+\frac{1}{N_0}\right)r^2}
\sum_{m,n}\frac{| m\rangle\langle n|}{\sqrt{m!n!}}
\nonumber\\
&&\times r^{m+n}e^{i(m-n)\theta}\left(1-\frac{C\bar{\nu}}{N_0}r^2\cos^2\theta+\frac{C\bar{\nu}}{2}\right)
\nonumber\\
&=& \hrho^{(0)}+\hrho^{(1)}~.
\eeqn
We find the steady state density matrix in the zeroth order of $\bar{\nu}$
\beqn
\hrho^{(0)}&=&\frac{1}{N_0}\sum_{m=0}^\infty\frac{| m\rangle\langle m|}{(1+1/N_0)^{m+1}}
\nonumber\\
&=&\frac{1}{Z_0}e^{-\beta\hbar\omega_0(\had\ha+1/2)}
\eeqn
where we use $1+1/N_0=e^{\beta\hbar\omega_0}$ and the zeroth order partition function $Z_0=N_0 e^{\beta\hbar\omega_0/2}$.
Using $\cos^2\theta=(e^{2i\theta}+e^{-2i\theta}+2)/4$, we find the correction in the first order of $\bar{\nu}$
\begin{widetext}
\beqn
\hrho^{(1)}&=&\frac{2C\bar{\nu}}{N_0}\sum_m\left[-\frac{1}{4N_0}\left\{\frac{|m\rangle\langle m+2|\sqrt{(m+2)(m+1)}}{(1+1/N_0)^{m+3}}
+\frac{|m\rangle\langle m-2|\sqrt{m(m-1)}}{(1+1/N_0)^{m+1}}\right\}\right.
\nonumber\\
&&~~~~~~~\left.-\frac{1}{2}|m\rangle\langle m|\left\{\frac{1}{N_0}\frac{m+1}{(1+1/N_0)^{m+2}}-\frac{1}{(1+1/N_0)^{m+1}}\right\}\right]
\nonumber\\
&=&-\frac{2C\bar{\nu}}{N_0}e^{-\beta\hbar\omega_0(\had\ha+1)}\left[\frac{1}{4N_0}\left(e^{-2\beta\hbar\omega_0}\ha\ha+\had\had\right)
+\frac{1}{2}\left\{\frac{1}{N_0}e^{-\beta\hbar\omega_0}(\had\ha+1)-1\right\}\right]
\nonumber\\
&=&-\hrho^{(0)}\frac{C\bar{\nu}}{2N_0}e^{-2\beta\hbar\omega_0}\left(\ha\ha+e^{2\beta\hbar\omega_0}\had\had+2e^{\beta\hbar\omega_0}\had\ha-2N_0e^{2\beta\hbar\omega_0}
\right)~.
\eeqn
\end{widetext}

\section{Dynamics}
S dissipates to steady state as $e^{-\mathsf{F}_{\rm eff} t}$. We get the eigenvalues of $\mathsf{F}_{\rm eff}$, the real part of eigenvalues can be found easily to be equal to $\omega_0 f_1/2$ to $\mathcal{O}(\bar{\nu})$. The relaxation rate (inverse relaxation time) is given by 
\begin{equation}
\tau_{\rm S}^{-1}= \omega_0f_1=\omega_0R_\mathrm{B}\frac{\bar{\nu}} {6}~,
\end{equation}
which is consistent with Eq.~(\ref{tau_S}) estimated from the total drift matrix $\mathsf{F}$, as expected. It is a typical property of dynamics dependent on thermostat. 

Time-dependent kernel $\mathsf{A}_\mathrm{S}(t)$ can be found from Eq.~(\ref{inverse_A}). We find $e^{-\mathsf{F}_{\rm eff} t}=e^{-t/\tau_{\rm S}}[e^{-i\omega_0't}|-\rangle\langle -|+e^{i\omega_0't}|+\rangle\langle +|]$ where $|\pm\rangle$ ($\langle\pm|$) denotes a right (left) eigenvector of $\mathsf{F}_{\rm eff}$. After some algebra, we find
\begin{widetext}
\begin{eqnarray}
\mathsf{A}(t)&=&\left(\begin{array}{cc}
\frac{2}{N_0\left[1-e^{-\frac{t}{\tau_{\rm S}}}\right]}+\frac{2 f_2N_0-8d_2+e^{-\frac{t}{\tau_{\rm S}}}\left[8d_2-2f_2 N_0\right]}{N_0^2\left[1-e^{-\frac{t}{\tau_{\rm S}}}\right]^2}&-\frac{e^{-\frac{t}{\tau_{\rm S}}}f_1N_0}{N_0^2\left[1-e^{-\frac{t}{\tau_{\rm S}}}\right]^2}
\\ -\frac{e^{-\frac{t}{\tau_{\rm S}}}f_1N_0}{N_0^2\left[1-e^{-\frac{t}{\tau_{\rm S}}}\right]^2}&
\frac{2}{N_0\left[1-e^{-\frac{t}{\tau_{\rm S}}}\right]}+\frac{e^{-\frac{t}{\tau_{\rm S}}}2d_2}{N_0^2\left[1-e^{-\frac{t}{\tau_{\rm S}}}\right]^2}
\end{array}\right)
\nonumber\\
&&+\left(\begin{array}{cc}
\frac{e^{-\frac{t}{\tau_{\rm S}}}\left[4 d_2\cos[2\omega_0' t]-f_1N_0\sin[2\omega_0't]\right]}{N_0^2\left[1-e^{-\frac{t}{\tau_{\rm S}}}\right]^2}&\frac{e^{-\frac{t}{\tau_{\rm S}}}\left[f_1N_0\cos[2\omega_0't]+4d_2\sin[2\omega_0't]\right]}{N_0^2\left[1-e^{-\frac{t}{\tau_{\rm S}}}\right]^2}
\\ \frac{e^{-\frac{t}{\tau_{\rm S}}}\left[f_1N_0\cos[2\omega_0't]+4d_2\sin[2\omega_0't]\right]}{N_0^2\left[1-e^{-\frac{t}{\tau_{\rm S}}}\right]^2}&
\frac{e^{-\frac{t}{\tau_{\rm S}}}\left[2d_2\cos[2\omega_0't]+f_1N_0\sin[2\omega_0't]\right]}{N_0^2\left[1-e^{-\frac{t}{\tau_{\rm S}}}\right]^2}~.
\end{array}\right)
\label{time_dependent_kernel}
\end{eqnarray}
\end{widetext}
One can observe that there appear $f_1$ and $\tau_S$ depending on the type of dissipator of B. 

As the system is initially isolated from the bath, we have $\rho_S(0)=\sum_{n,m}\rho_{nm}|n\rangle\langle m|$. The corresponding $P$-distribution is given as 
\begin{equation}
P_{\rm S}(z_0, 0)=\sum_{n,m}\rho_{nm}\frac{e^{|z_0|^2}}{\sqrt{n!m!}}\frac{\partial^{n+m}}{\partial(-z_0)^n\partial(-\bar{z}_0)^m}\delta^{(2)}(z_0)~
\end{equation}
which is not positive and highly singular, as well known~\cite{glauber1963,sudarshan1963}. As a simple case, we consider $\rho_{\rm S}(0)=|1\rangle\langle 1|$. Then, we can get the marginal $P$-distribution at time $t$ by $\int d^{2}z_0 P_{\rm S}(z,t|z_0,0)P_{\rm S}(z_0,0)$. Representing by $\mathbf{r}$ and integrating by parts, we obtain 
\begin{eqnarray}
\lefteqn{P_{\rm S}(\mathbf{r},t)=
\frac{[{\rm det}\mathsf{A}(t)]^{1/2}}{8\pi} \exp\left[-\frac{1}{2}\mathbf{r}^{\rm t}\cdot\mathsf{A}(t)\cdot\mathbf{r}\right]}
\nonumber\\
&&~~~~~~~~~\times\Big[4-{\rm Tr}\hat{\mathsf{K}}(t)+\left(x\mathsf{K}(t)_{11}+y\mathsf{K}(t)_{21}\right)^2\nonumber\\
&&~~~~~~~~~~~~~~~+\left(x\mathsf{K}(t)_{12}+y\mathsf{K}(t)_{22}\right)^2\Big]
\end{eqnarray}
where $\mathsf{K}(t)=\mathsf{A}(t)e^{-\mathsf{F}_{\rm eff}t}$, $\hat{\mathsf{K}}(t)=^{-\mathsf{F}_{\rm eff}^{\rm t}t}\mathsf{A}(t)e^{-\mathsf{F}_{\rm eff}t}$,  and subscripts denote matrix elements. It has negative distribution for an intermediate period from the beginning and becomes positive everywhere afterwards, approaching the steady state distribution. We show the distributions for various times in Figs.~1 and 2 in the text. 

\bibliography{bib}